\documentclass[12pt]{article}
\usepackage{graphicx,tabularx,epsfig}
\usepackage{color}
\usepackage{enumitem}
\usepackage{caption}
\usepackage{subcaption}
\usepackage{amsmath}
\usepackage{amssymb}
\usepackage{amsthm}
\usepackage{physics}
\usepackage{dsfont, mathtools}
\usepackage{psvectorian}
\usepackage{blkarray}
\usepackage{bm}
\usepackage{url}
\usepackage{setspace}

\usepackage{xcolor}
\colorlet{linkequation}{blue}

\usepackage[colorlinks = true,
            linkcolor = blue,
            urlcolor  = blue,
            citecolor = blue,
            anchorcolor = blue]{hyperref}

\usepackage{floatrow}

\usepackage{amsthm}

\usepackage{amssymb}

\newlength{\abstractwidth}
\renewcommand{\thefootnote}{\fnsymbol{footnote}}
\renewcommand{\thanks}[1]{\footnote{#1}} 
\newcommand{\starttext}{
\setcounter{footnote}{0}
\renewcommand{\thefootnote}{\arabic{footnote}}}

\makeatletter
\g@addto@macro\normalsize{%
  \setlength\abovedisplayskip{10pt}
  \setlength\belowdisplayskip{20pt}
  \setlength\abovedisplayshortskip{10pt}
  \setlength\belowdisplayshortskip{20pt}
}
\makeatother

\usepackage{color}

\renewcommand{\title}[1]{\vbox{\center\LARGE{#1}}\vspace{5mm}}
\renewcommand{\author}[1]{\vbox{\center#1}\vspace{5mm}}
\newcommand{\address}[1]{\vbox{\center\em#1}}
\newcommand{\email}[1]{\vbox{\center\tt#1}\vspace{5mm}}

\theoremstyle{definition}

\begin{document}

\singlespacing

\begin{titlepage}
\rightline{}
\bigskip
\bigskip\bigskip\bigskip\bigskip
\bigskip
\centerline{\Large \bf {Uncomplexity and Black Hole Geometry}}

\bigskip \noindent

\bigskip
\begin{center}


\author{Ying Zhao}

\address{Stanford Institute for Theoretical Physics and Department of Physics, \\
Stanford University, Stanford, CA 94305-4060, USA }

\email{zhaoying@stanford.edu}

\bigskip

\end{center}

\begin{abstract}

We give a definition of uncomplexity of a mixed state without invoking any particular definitions of mixed state complexity, and argue that it gives the amount of computational power Bob has when he only has access to part of a system. We find geometric meanings of our definition in various black hole examples, and make a connection with subregion duality. We show that Bob's uncomplexity is the portion of his accessible interior spacetime inside his entanglement wedge. This solves a puzzle we encountered about the uncomplexity of thermofield double state. In this process, we identify different kinds of operations Bob can do as being responsible for the growth of different parts of spacetime.

\medskip
\noindent
\end{abstract}

\end{titlepage}

\starttext \baselineskip=17.63pt \setcounter{footnote}{0}

{\hypersetup{hidelinks}
\tableofcontents
}

\setcounter{equation}{0}
\section{Introduction}\label{Introduction}

It was pointed out in \cite{Brown:2017jil} that one can do useful computations with a system away from being maximally complex. In \cite{Brown:2017jil} the concept of uncomplexity was introduced to characterize the amount of computational power one has with a system in some particular state. The uncomplexity of a pure state is defined as the difference of the maximal complexity and the state complexity. 

For a strongly interacting holographic system in a state with a dual black hole geometry, it was conjectured that its state complexity will increase linearly with time until it saturates at some maximal value, $\sim e^S$ for a system with coarsed grained entropy $S$ \cite{Susskind:2015toa}\cite{Susskind:2017talk}. At any time before that, the state has some computational power given by its uncomplexity.

On the other hand, for a state with a classical gravity dual, the growth of the wormhole was conjectured to reflect the increase of the state complexity \cite{Stanford:2014jda}\cite{Roberts:2014isa}\cite{Brown:2015bva}\cite{Brown:2015lvg}. It was also conjectured that a black hole state with increasing complexity has a smooth horizon \cite{Susskind:2015toa}. The authors of \cite{Brown:2017jil} made the connection that the uncomplexity of a black hole state corresponds to the interior spacetime acessible for an infalling observer to jump in. 

In the first part of the paper, we generalize the concept of uncomplexity to mixed states. Following the simple assumption that the computational power of a mixed state does not depend on its purifications, we give a definition of the uncomplexity of a density matrix without invoking any definitions of complexity of a mixed state. In this process we study different kinds of operations Bob can apply if he only has access to a subsystem of an entangled state. 

\begin{figure}[H] 
\label{onesideBH}
 \begin{center}                      
      \includegraphics[width=2.6in]{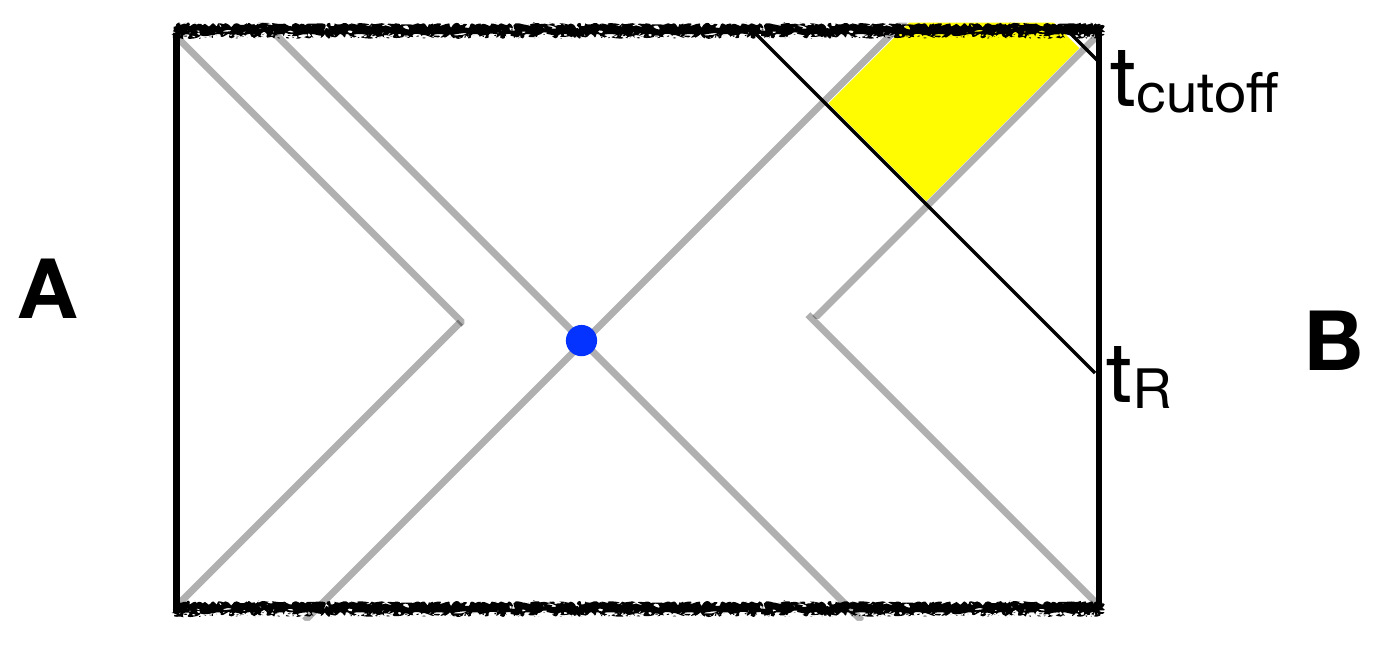}
      \caption{The blue dot is the entangling surface between A and B.}
  \label{uncomplexityspacetime_intro}
  \end{center}
\end{figure}

We then discuss the connection between uncomplexity and the black hole geometry. With our definition, we give geometric interpretations of uncomplexity in various examples, and show how it fits together with subregion duality \cite{Dong:2016eik}. We see that the uncomplexity of a density matrix exactly corresponds to the portion of interior spacetime accessible to an observer which is also inside the entanglement wedge of the density matrix (Figure \ref{uncomplexityspacetime_intro}). In this process we show that the different kinds of operations we studied earlier are responsible for the growth of different spacetime regions. We give a quantum circuit description of a black hole with a wide Penrose diagram whose entangling surface is behind its event horizon. We also point out the role played by the apparent horizon \cite{Engelhardt:2017aux}. 

\section{Uncomplexity of a density matrix}

A maximally complex state has no computational power. In analogy to thermodynamics, the second law of complexity states that the complexity of a state always increases. It takes much fine-tuning to decrease the complexity \cite{Brown:2017jil}. A state that is not maximally complex is like a thermodynamic system away from its equilibrium. One can do useful computations with a state that is not maximally complex. In analogy to the free energy as the amount of energy that can be used to do work, uncomplexity characterizes the computational power of a state. The uncomplexity of a pure state was defined as the difference between the maximal complexity and the complexity of the state \cite{Brown:2017jil}\cite{Susskind:2017talk}.
\begin{align*}
\mathcal{UC}(\ket{\psi})\coloneqq \mathcal{C}_{\text{max}} -\mathcal{C}(\ket{\psi})
\end{align*}

We will generalize this concept to a mixed state. There have been various attempts to define the complexity of a density matrix \cite{Swingle:2017talk}. It's complicated by the fact that the density matrix has different purifications. Here, we will emphasize that  it's not necessary to have a definition of mixed state complexity in order to define the uncomplexity of a density matrix. This follows from the simple assumption that the uncomplexity of a density matrix does not depend on its purifications. The question we want to answer is, what's the computational power of a density matrix $\rho_B$? Alternatively, we can ask this question in the following way: Imagine there is a system AB in a pure state $\ket{\psi} \in \mathcal{H}_A\otimes\mathcal{H}_B$, where $\ket{\psi}$ is an arbitrary purification of $\rho_B$: $\text{tr}_A\left(\ket{\psi}\bra{\psi}\right) = \rho_B$. How much computational power can Bob get out of the state $\ket{\psi}$ when he only has access to the subsystem B?

Bob can apply arbitrary unitary operators $U_B$ to his subsystem: $\ket{\psi}\rightarrow U_B\ket{\psi}$. Without fine-tuning, generically $U_B$ will increase the complexity of the state $\ket{\psi}$. As a first try, we let the uncomplexity of the density matrix $\rho_B$ be
\begin{align*}
\mathcal{UC}(\rho_B)\stackrel{?}{\coloneqq} \max_{U_B\text{ unitary on B}}\mathcal{C}(U_B\ket{\psi})-\mathcal{C}(\ket{\psi})\ \ \ \text{INCORRECT}
\end{align*}
This is incorrect. The reason is, not all $U_B$'s that increase the complexity of $\ket{\psi}$ can be useful computations for Bob. Imagine a state $\ket{\psi}$ where Bob and Alice share some Bell pairs. Subsystem B is maximally mixed, i.e. $\rho_B$ is proportional to the  identity matrix. Bob can apply unitaries $U_B$ to increase the complexity of state $\ket{\psi}$, but his density matrix $\rho_B$ will stay the same. In other words, even though the state $\ket{\psi}$ changes, Bob is not able to see it. To exclude such $U_B$'s (which do not change Bob's density matrix), we use the following definition:
\begin{align}
\label{uncomplexitydef}
\mathcal{UC}(\rho_B) \coloneqq\max_{U_B \text{ unitary on B}}\mathcal{C}(U_B\ket{\psi})-\max_{U_B\  \text{does not change}\  \rho_B}\mathcal{C}(U_B\ket{\psi}).
\end{align}

 In the first term of \eqref{uncomplexitydef}, $U_B$  is taken from all unitaries Bob can apply to his subsystem B without fine tuning. In the second term, we exclude those operators which do not change his density matrix. To further motivate this definition, we need to better understand the second term in \eqref{uncomplexitydef}. 

\textbf{Claim}: If an unitary operator $U_B$ does not change the density matrix $\rho_B$, then $U_B$ can be undone by some unitary operator on A, i.e. there exists $U_A$, s.t. $U_B\ket{\psi} = U_A\ket{\psi}$. 

To see this, we go to Schmidt basis. Bob's density matrix can be written as 
\begin{equation}
\label{density_matrix}
\begin{blockarray}{ccccccc}
 \begin{block}{c(ccccc)c}
&\lambda_1\mathds{1} &0 & \dots &0 & 0 & \bm{\bigg\}}N_1\\
&0 &\lambda_2\mathds{1}& & 0 & 0& \bigg\}N_2 \\
\rho_B = &\vdots& &\ddots & &\vdots&\vdots\\
&0&0 & & \lambda_k\mathds{1}&0&\bigg\}N_k\\
&0&0 & \dots& 0& 0& \bigg\}N_0\\
\end{block}
&\vspace{-.9cm}&&&&&\\
&\underbrace{}&\underbrace{}& &\underbrace{}&\underbrace{}&\\
&N_1 & N_2 &\dots & N_k & N_0
\end{blockarray}
\end{equation}

\begin{figure}[H]
 \begin{center}
  \begin{subfigure}[b]{0.4\textwidth}
  \begin{center}
    \includegraphics[scale=0.5]{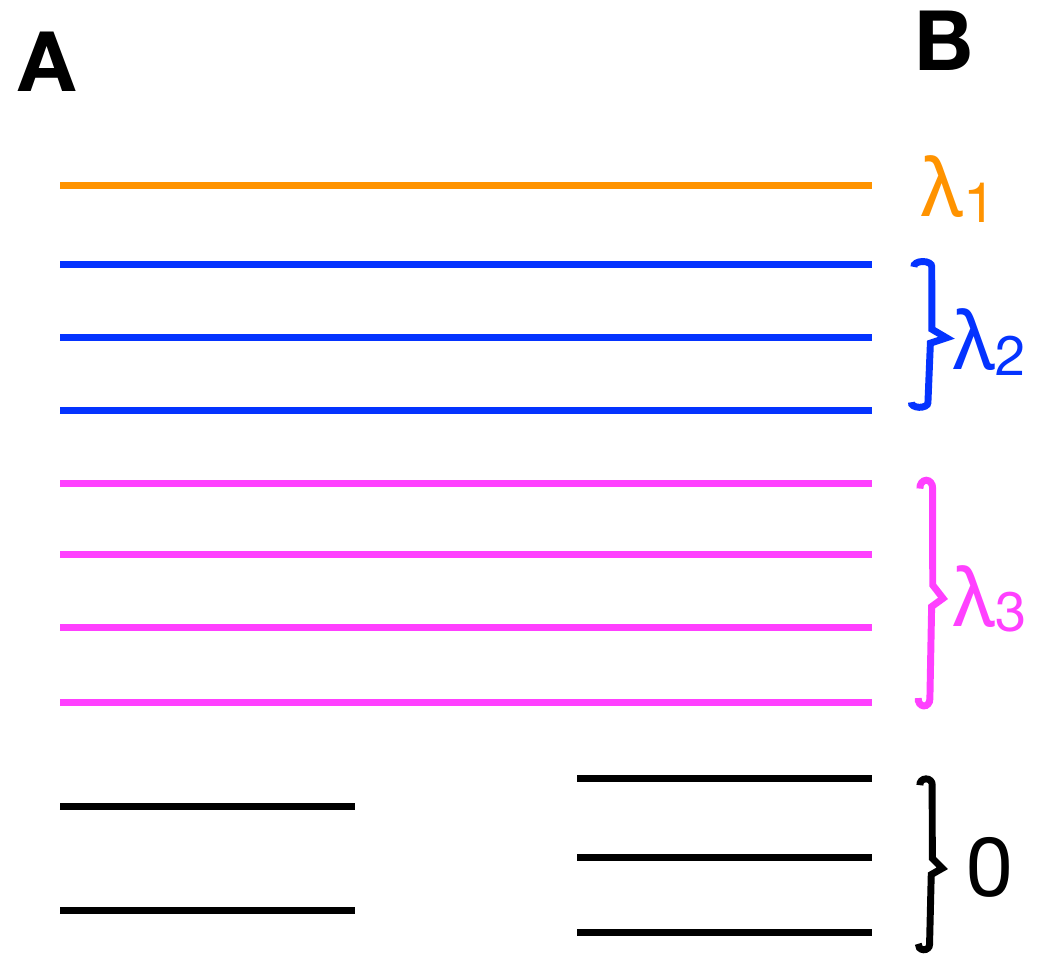}
    \caption{}
    \label{schmidt1}
    \end{center}
  \end{subfigure}
  \hspace{2em}
  \begin{subfigure}[b]{0.4\textwidth}
  \begin{center}
    \includegraphics[scale=0.5]{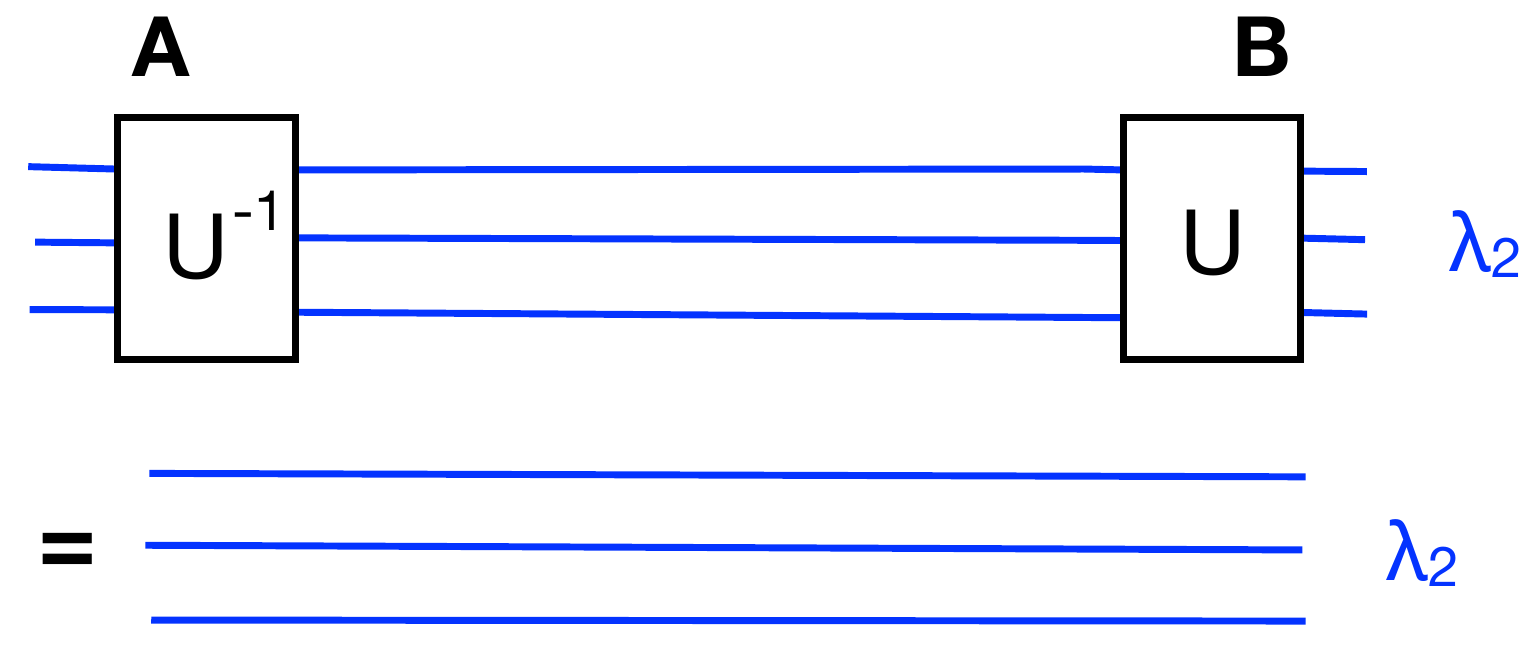}
    \caption{}
    \label{schmidt2}
    \end{center}
  \end{subfigure}
  \caption{A pictorial representation of the density matrix in Schmidt basis. The lines here represent the Schmidt eigenbasis, not qubits. }
  \vspace{-.5cm}
  \label{schmidt}
  \end{center}
\end{figure}

In \eqref{density_matrix}, we assume $\rho_B$ has $k$ nonzero distinct eigenvalues, $\lambda_i$ with degeneracy $N_i$, $i = 1, ..., k$.  The rest $N_0$ eigenvalues are zero, corresponding to eigenvectors not entangled with A system. Figure \ref{schmidt1} is a pictorial representation of \eqref{density_matrix}. $U_B$'s that do not change $\rho_B$ are unitary rotations on each identity block, i.e. $SU(N_1)\times...\times SU(N_k)\times SU(N_0)$. Rotations from $SU(N_i), \ i = 1,..., k$ can be undone from A system (Figure \ref{schmidt2}), while rotations from $SU(N_0)$ do not change the state $\ket{\psi}$. 

We see that if Bob has a subsystem of an entangled state, there are two different kinds of unitary operators he can apply: those that can be undone from the other side, and those that cannot. The first kind are relative rotations of the Schmidt basis. They are really two-sided operations and do not belong to any single density matrix. One can only see their effect if one has control of both subsystems, and Bob alone cannot do computations with them. The second kind of operations mix eigenvectors belonging to different Schmidt eigenvalues. They will change Bob's density matrix, and Bob can do useful computations with them. 

Now imagine for our AB system, Alice has applied some $U_A$ to make the state $\ket{\psi}$ as complex as possible. If Bob does those relative Schmidt basis rotations, the state complexity can only decrease or stay the same.\footnote{Bob's relative Schmidt basis rotations will depend on the state. See Appendix \ref{uncomplexity_careful_treatment} for more careful discussions.} Any further complexity increase from Bob's operations must come from rotations of eigenvectors belonging to different Schmidt eigenvalues. For such a state, \eqref{uncomplexitydef} gives Bob's computational power, so we can consider it as the uncomplexity of Bob's density matrix. 

Here, we motivated definition \eqref{uncomplexitydef}, but didn't give the most rigorous treatment. A more careful discussion is given in Appendix \ref{uncomplexity_careful_treatment}. We'll use \eqref{uncomplexitydef} in the rest of the paper. Our focus will be on black hole geometries. For reasons that will be explained later, \eqref{uncomplexitydef} has a particularly clear circuit picture as well as geometric meaning in case of black holes.

\section{Black hole geometry interpretations}

In this section, we'll apply \eqref{uncomplexitydef} to the case of black holes, and illustrate the idea that the growth of spacetime is fueled by uncomplexity. We'll also see how it fits together with subregion duality \cite{Dong:2016eik}.

\subsection{Simple examples}
\label{simple_examples}

$\bullet$ One-sided black hole

Start from empty $AdS$ space. At $t = 0$, we send in a spherical shell of matter from the boundary, and a black hole forms. After that, the interior starts to grow. This growth corresponds to the increase of state complexity. Equivalently, one can say that the growth is fueled by the computational power, i.e. uncomplexity of the state. At a very late time $t_{\text{cutoff}}$, the growth will stop when there is no more uncomplexity to exploit. At any time $t$, the uncomplexity left corresponds to the potential for the spacetime to grow, i.e. the interior spacetime left that someone can safely jump in. It decreases linearly with time.\footnote{The interior spacetime acessible to Bob shrinks as time increases. We can quantify the spacetime with different quantities, e.g. volume of maximal surface \cite{Stanford:2014jda}\cite{Roberts:2014isa}, action \cite{Brown:2015bva}\cite{Brown:2015lvg}, spacetime volume, and so on. All these quantities decrease linearly with time. It's a result of the Rindler boost symmetry across the horizon \cite{Zhao:2017iul}.} Figure \ref{onesideBH} is the geometric interpretation of \eqref{uncomplexitydef} in this case.

\begin{figure}[H] 
\label{onesideBH}
 \begin{center}                      
      \includegraphics[width=4.5in]{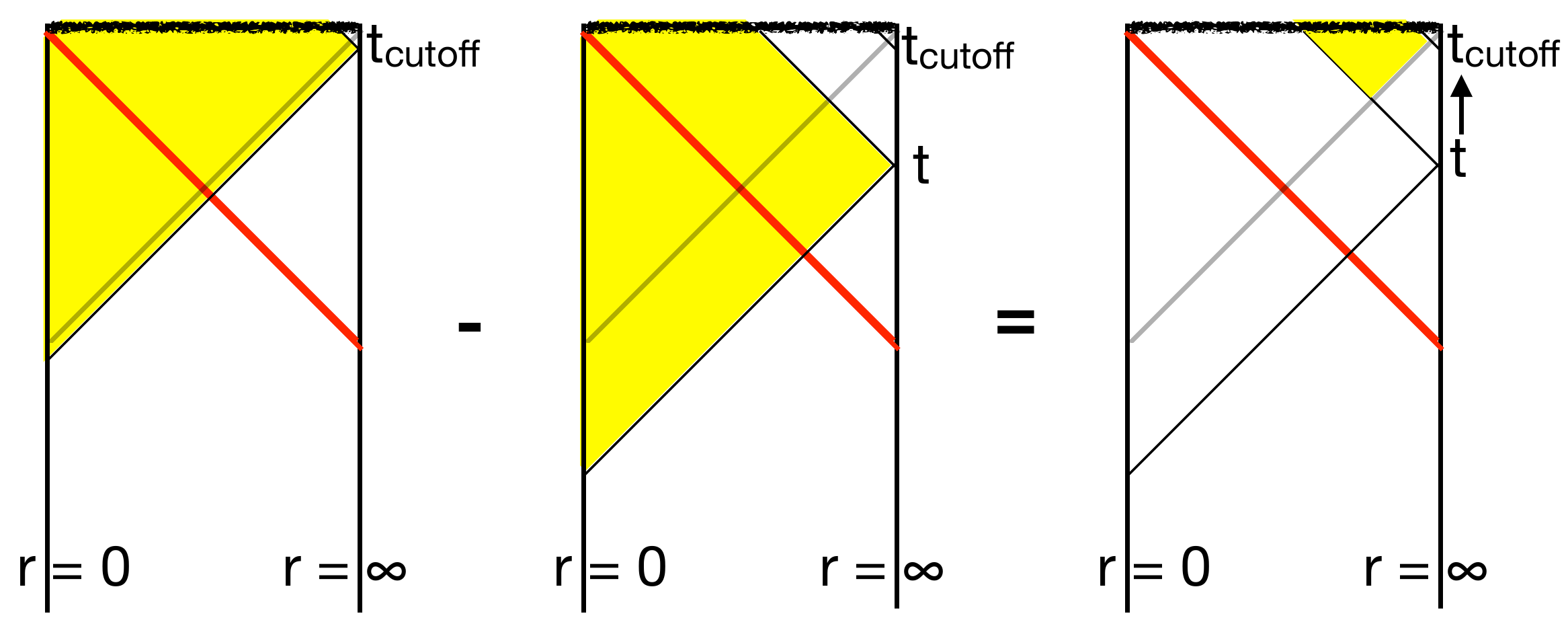}
      \caption{}
  \label{onesideBH}
  \end{center}
\end{figure}

\ \\
$\bullet$ Thermofield Double (Two-sided black hole)

\begin{figure}[H]
 \begin{center}
  \begin{subfigure}[b]{0.4\textwidth}
  \begin{center}
    \includegraphics[scale=0.5]{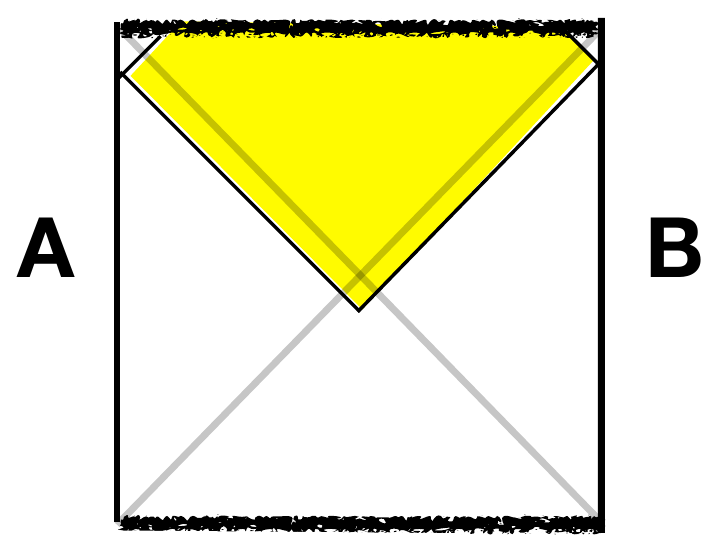}
    \caption{}
    \label{thermofield_double}
    \end{center}
  \end{subfigure}
  \hspace{2em}
  \begin{subfigure}[b]{0.4\textwidth}
  \begin{center}
    \includegraphics[scale=0.5]{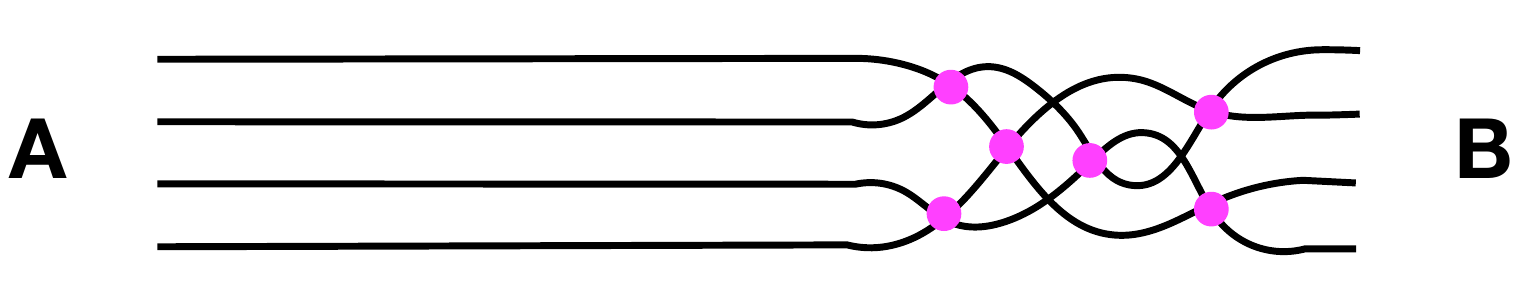}
    \caption{The lines here represent both qubits and Schmidt eigenbasis.}
    \label{thermofield_double_circuit}
    \end{center}
  \end{subfigure}
  \caption{}
  \vspace{-.5cm}
  \label{schmidt}
  \end{center}
\end{figure}

Next, we look at the unperturbed thermofield double. We call the left black hole subsystem A (Alice' side), and the right black hole subsystem B (Bob's side). We focus on subsystem B. Note that system B has coarsed grained entropy $S$, while its entanglement entropy with system A is also $S$. If we focus on the near horizon region where the black hole is doing computations, all degrees of freedom are entangled with the other side. In a simplified model, we represent such a two-sided black hole by $S$ bell pairs plus a MERA-like RG circuit lying outside the horizon \cite{Swingle:2012wq}\cite{Hartman:2013qma}\cite{Brown:2015lvg}. The MERA circuit is time-independent, and only the $S$ qubits near horizon are doing computations. In this case, unitary operators on B do not change the density matrix. So the two terms appearing in \eqref{uncomplexitydef} are the same and correspond to a wormhole with maximal length (Figure \ref{thermofield_double}).  The uncomplexity of subsystem B is zero, which is consistent with the expectation that a completely mixed density matrix has no computational power.

At first sight this result looks a bit confusing. We know the wormhole can get very long, and there is a lot of interior spacetime for Bob to jump into. How is this consistent with the fact that the uncomplexity of Bob's subsystem is zero? This paradox will be resolved in this paper. From Figure \ref{thermofield_double_circuit}, we see that the unitary operations from either side give relative rotations of the Schmidt basis. These relative rotations can make the overall circuit long and complex, but one cannot say which side these operations belong to. This is the boost symmetry of thermofield double. If we look at the Penrose diagram (Figure \ref{thermofield_double}), the long wormhole lies outside the entanglement wedge of both CFT's. Even though Bob can jump into the interior, with full knowledge of right CFT one still doesn't know Bob's experience once he crosses the horizon. Alice can send him a flower or bullet.\footnote{I heard this saying from Leonard Susskind.} Neither Bob nor Alice has full control of the interior which really belongs to both sides. The existence of the interior spacetime comes from the entanglement of the two sides. This is ER = EPR \cite{Maldacena:2013xja}, and is very mysterious. 

\begin{figure}[H]
 \begin{center}
 \hspace{-3cm}
  \begin{subfigure}[b]{0.4\textwidth}
   \hspace{-3cm}
  \begin{center}
  \hspace{-1.6cm}
    \includegraphics[scale=0.5]{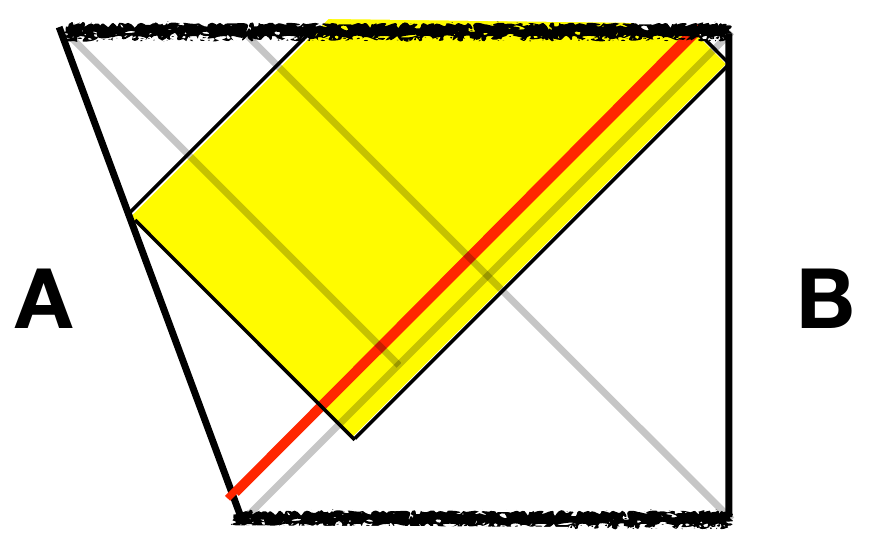}
    \caption{}
    \label{thermofield_double_A}
    \end{center}
  \end{subfigure}
  \hspace{-1em}
  \begin{subfigure}[b]{0.4\textwidth}
  \begin{center}
    \includegraphics[scale=0.5]{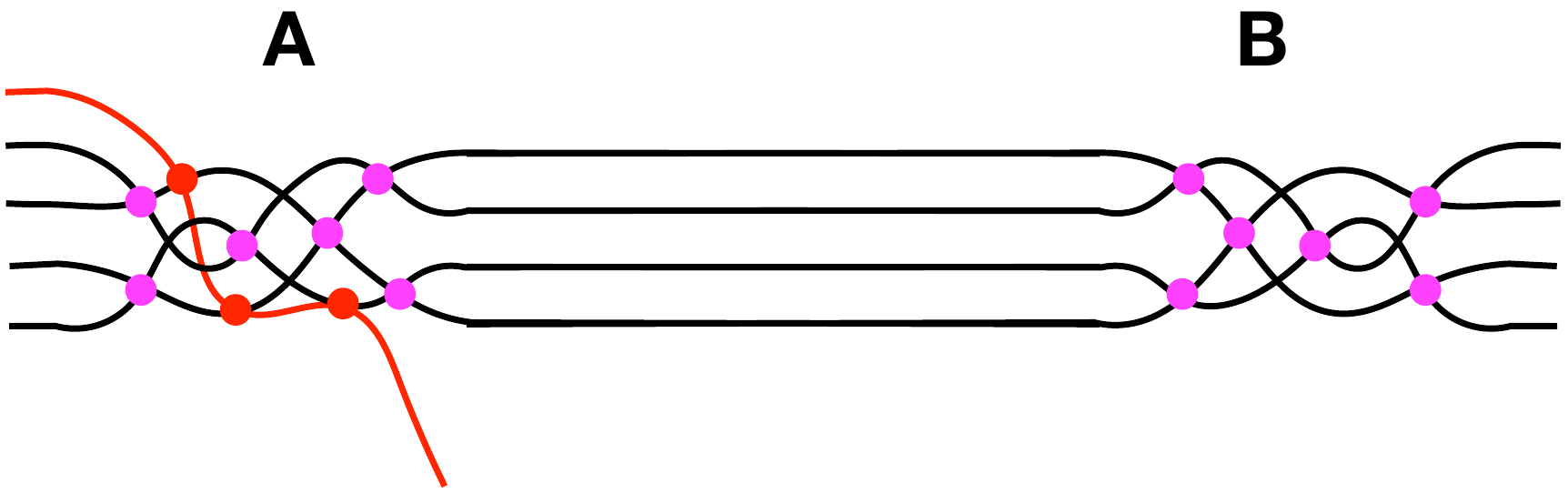}
    \caption{The lines here represent qubits. Alice' extra photon is in red.}
    \label{thermofield_double_circuit_A}
    \end{center}
  \end{subfigure}
  \caption{}
  \vspace{-.5cm}
  \label{schmidt}
  \end{center}
\end{figure}

We can also look at thermofield double perturbed by Alice throwing in a thermal photon. After the extra photon gets scrambled with her original $S$ qubits, it will not be easy for Alice to undo Bob's operations. But nevertheless Bob's unitary operations can still be undone by Alice (Figure \ref{thermofield_double_circuit_A}). In this case, to maximize both terms in \eqref{uncomplexitydef}, we can put Bob's time at some large cutoff time (Figure \ref{thermofield_double_A}), and the two terms will still be equal. Bob's density matrix still has no uncomplexity as expected. In Figure \ref{thermofield_double_A}, the interior is outside Bob's entanglement wedge.

\subsection{Transition from two-sided to one-sided black holes}
\label{transition}

One key feature of a one-sided black hole is that in principle Bob can predict an infalling observer's experience within the framework of quantum mechanics.\footnote{It may be highly complex.} In other words, he has full control of his interior. From a quantum information point of view, the expansion of the interior is fueled by Bob's own uncomplexity. In contrast, if Bob holds one side of thermofield double, he does not have full control of the interior since the expansion of the interior is fueled by the relative rotations of the Schmidt basis, not Bob's own uncomplexity. In the rest of the paper,  by saying one-sided black hole, we mean that the portion of the interior accessible to Bob is determined only by operations on his subsystem. By saying two-sided black hole, we mean the portion of the interior accessible to Bob can be also affected by operations on Alice' subsystem. Note that the property of being a one-sided or two-sided black hole can be time-dependent. We DO NOT mean counting the number of boundaries the system has.\footnote{For example, a very wide Penrose diagram for which the entangling surface is deep inside Bob's event horizon is considered as a one-sided black hole for Bob.  }

Next we will perturb thermofield double from Bob's side, and ask for Bob's uncomplexity. In this case, we expect Bob to have finite uncomplexity after the perturbation. See \cite{Brown:2017jil}\cite{Susskind:2017talk} for discussions of one clean qubit computation.\footnote{The original version of one clean qubit computaiton made use of a maximally mixed state \cite{Knill:1998wi}, not a maximally complex state as in \cite{Brown:2017jil}. This paper provides an uncomplexity definition in the context of \cite{Knill:1998wi}. }  We throw in an extra photon from Bob's side at time $t_w$.

\subsubsection{Quantum circuit piture}
\begin{figure}[H] 
 \begin{center}                      
      \includegraphics[width=4in]{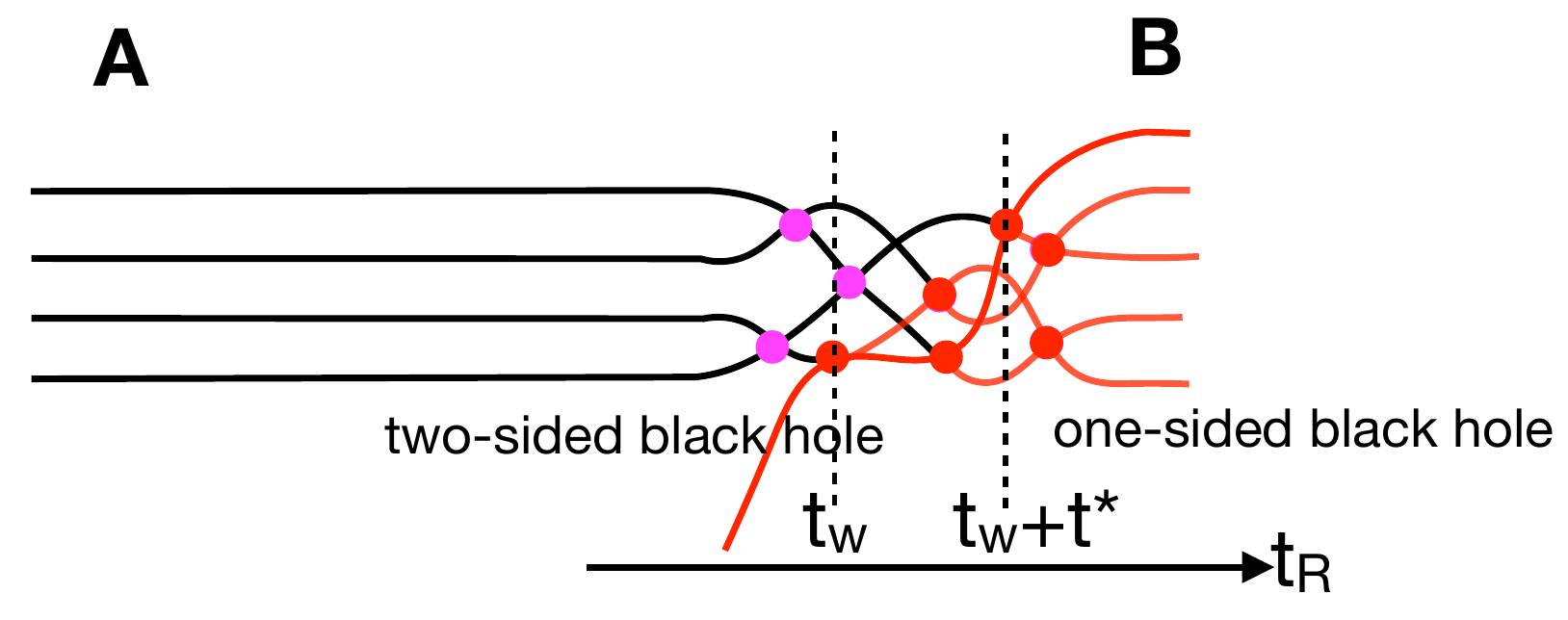}
      \caption{The lines here are qubits. The extra photon comes in at time $t_w$ as a red line.}
  \label{thermofield_double_circuit_B}
  \end{center}
\end{figure}

 Let's look at this from a quantum circuit point of view (Figure \ref{thermofield_double_circuit_B}). In Figure \ref{thermofield_double_circuit_B}, the qubits turn red after they interact directly or indirectly with the extra qubit. Before the extra photon comes in ($t_R<t_w$), Bob holds S qubits maximally entangled with Alice' qubits and he has no computational power. He has one side of a two-sided black hole. The extra photon brings him uncomplexity of order $e^S$ \cite{Brown:2017jil}\cite{Susskind:2017talk}. When $t_w<t_R<t_*$, the effect of the extra photon starts to spread. Since the number of qubits affected by the extra photon grows exponentially with time \cite{Susskind:2014jwa}\cite{Roberts:2014isa}\cite{Brown:2016wib}, during this regime most of the qubits have not been affected, and most of Bob's operations are still relative Shimidt basis rotations. So to a first approximation,\footnote{We'll do better in the next section.} in this regime we will ignore the interactions involving the extra photon and treat the system as $S$ maximally entangled pairs with one clean qubit in Bob's side. The uncomplexity of his density matrix does not change during this period. After the extra photon gets completely scrambled ($t_R>t_w+t_*$), all qubits are affected by the extra photon and unitary operators on Bob's side can no longer be undone from Alice side. We could say Bob has a one-sided black hole. His uncomplexity decreases linearly with time. 

\subsubsection{Black hole geometry}

Now we look at the black hole geometry. The extra photon comes in from Bob's side at time $t_w$. The transition from two-sided to one-sided black holes for Bob happens between $t_w$ and $t_w+t_*$. For less than scrambling time after we throw in the extra photon, the perturbation has little effect on the black hole geometry and we will ignore it in this section. We'll treat the geometry as an unperturbed two-sided black hole for $t_R<t_w+t_*$. We'll do better in the next section.

First, observe that in maximizing both terms in \eqref{uncomplexitydef}, we are free to move the left time, since the left time evolution can be written as some unitary operator on the right system (system B) which does not change Bob's density matrix. To maximize both terms in \eqref{uncomplexitydef}, we can fix the left time at either some large positive cutoff time, or some large negative cutoff time. Both will work. Here, we also see why we choose to work with definition \eqref{uncomplexitydef}. Black hole dynamics is expected to be chaotic and naturally push the state towards becoming maximally complex \cite{Susskind:2015toa}, so both terms in \eqref{uncomplexitydef} have geometric interpretations in the Penrose diagram. \\

\begin{figure}[H] 
 \begin{center}                      
      \includegraphics[width=5in]{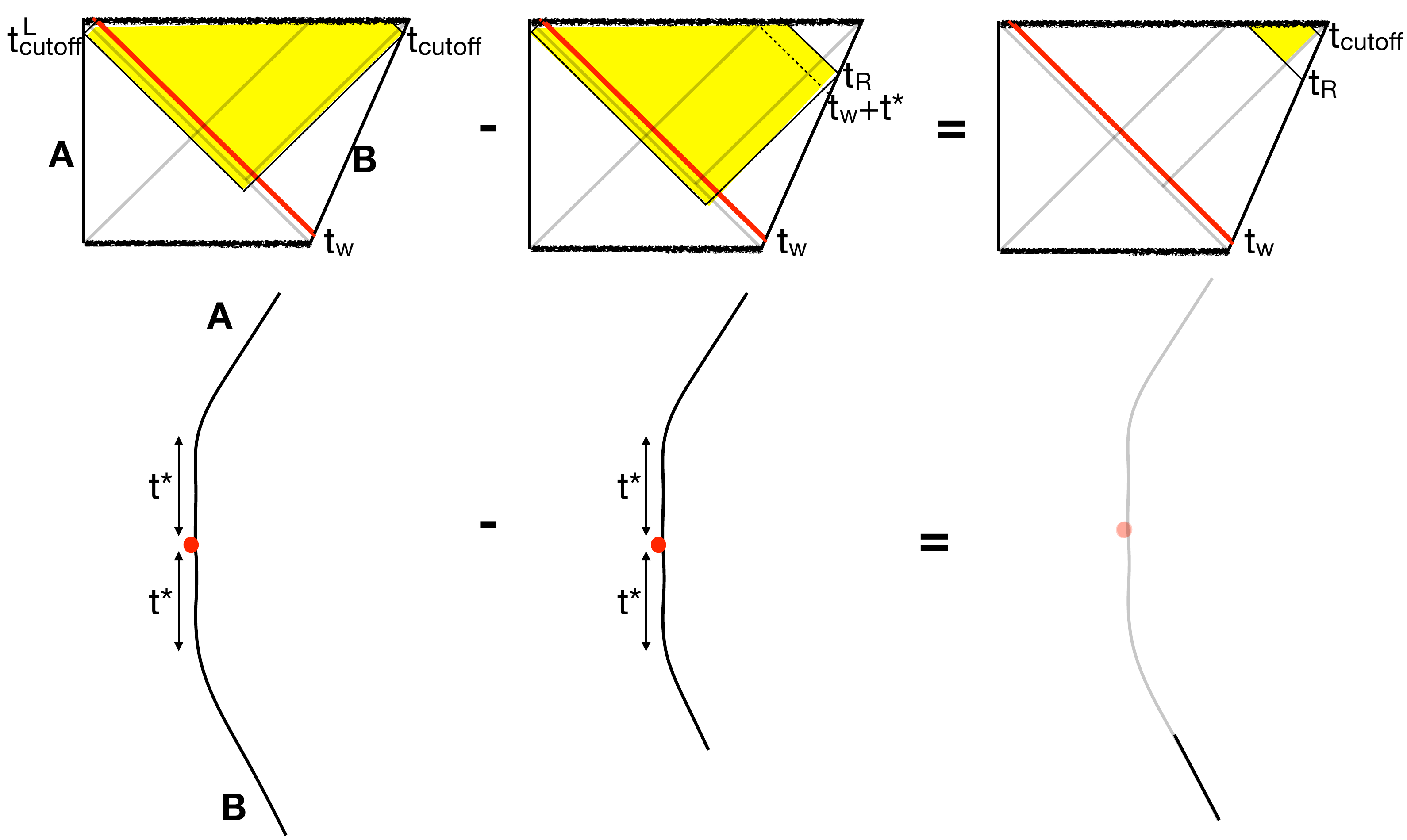}
      \caption{The pictures below the Penrose diagrams represent quantum circuits. See \cite{Stanford:2014jda}\cite{Brown:2016wib} for more details. }
  \label{later_than_scrambling_time_B2}
  \end{center} 
\end{figure}

We fix the left time at some large positive cutoff. We start by putting the right time at some large cutoff value $t_{\text{cutoff}}$ where Bob has no more uncomplexity. During the time when $t_w+t_*<t_R<t_{\text{cutoff}}$, the uncomplexity of Bob's density matrix decreases linearly with increasing right time (Figure \ref{later_than_scrambling_time_B2}),\footnote{In contrast to the case of unperturbed thermofied double, the left cutoff time does not move as we vary the right time due to the existence of the perturbation.} as the expansion of his black hole consumes his uncomplexity. In this regime, Bob essentially has a one-sided black hole. 

\begin{figure}[H] 
 \begin{center}                      
      \includegraphics[width=5in]{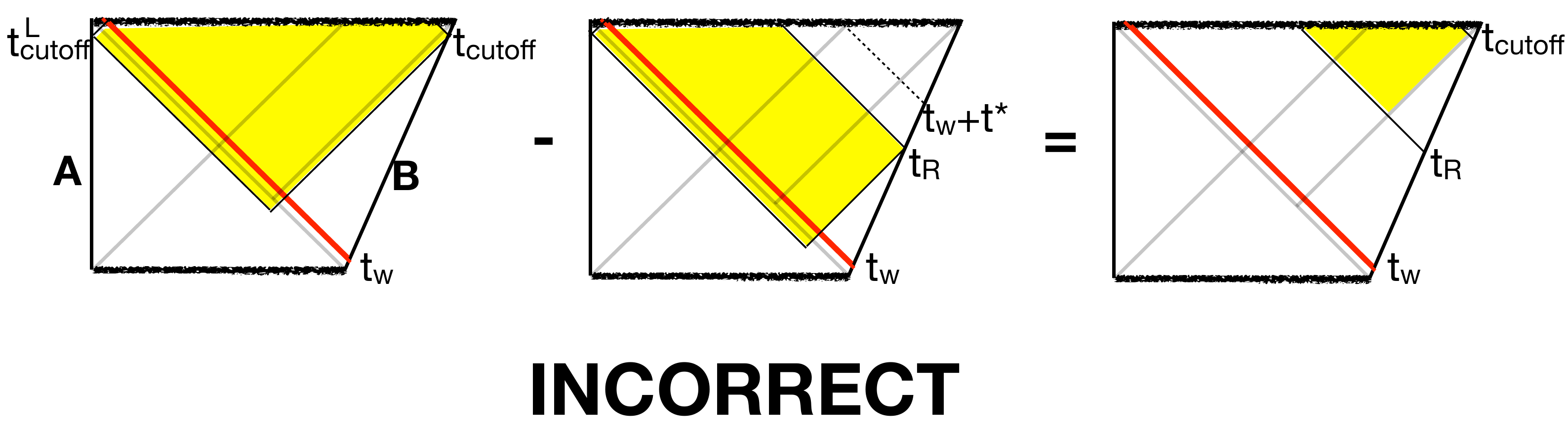}
      \caption{}
  \label{earlier_than_scrambling_time_B2INCORRECT}
  \end{center}
  \vspace{-0.5cm}
\end{figure}

When $t_w<t_R<t_w+t_*$, naively Bob's uncomplexity keeps increasing linearly with decreasing $t_R$ (Figure \ref{earlier_than_scrambling_time_B2INCORRECT}). However, from earlier circuit analysis we know this is incorrect. 

\begin{figure}[H] 
 \begin{center}                      
      \includegraphics[width=5in]{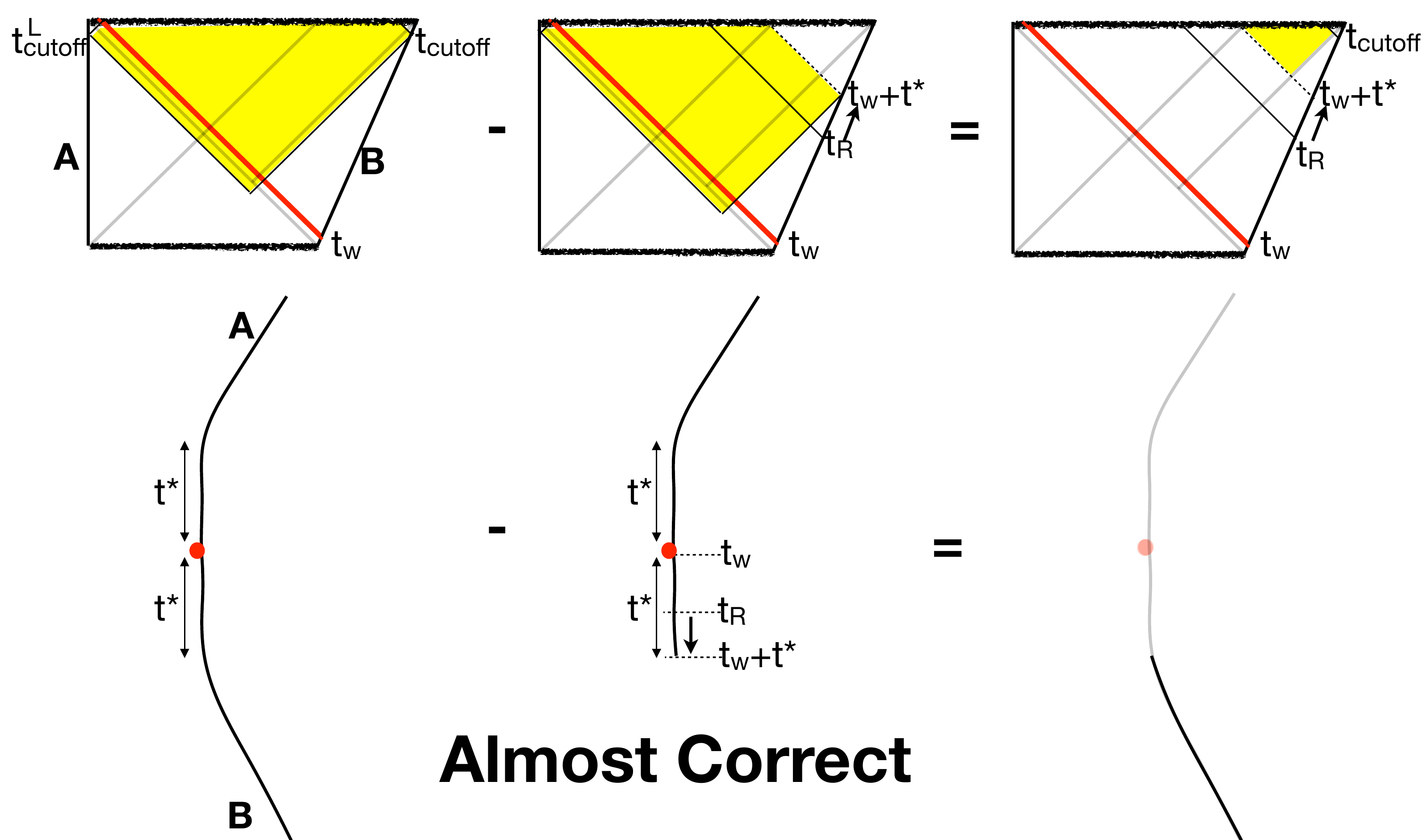}
      \caption{}
  \label{earlier_than_scrambling_time_B2CORRECT}
  \end{center}
\end{figure}

The reason is that in the second term of \eqref{uncomplexitydef}, we should maximize the uncomplexity of the state by Bob's unitary operations with the restriction that his density matrix stays the same. When we are at less than scrambling time after the perturbation, we could ignore the perturbation. We are essentially in the two-sided black hole regime, and the right time evolution until $t_w+t_*$ won't change the density matrix. So an almost correct picture is in Figure \ref{earlier_than_scrambling_time_B2CORRECT}. In this regime the expansion of the interior is fueled by the relative rotations of the Schmidt basis, and the uncomplexity of Bob's density matrix stays constant.

We can do the same calculation with the left cutoff time fixed at some large negative value and we'll get the same results. The details are in Appendix \ref{othercutoff}. \\

Notice a couple of points: 
\begin{itemize}
\item
The spacetime region corresponding to the uncomplexity of Bob's density matrix always stays inside Bob's entanglement wedge. See the Penrose diagrams on the right side of Figure \ref{later_than_scrambling_time_B2}, \ref{earlier_than_scrambling_time_B2CORRECT}. 
\item 
In the discussions of both the circuit picture and the black hole geometry, we made the approximation that Bob's density matrix does not change by right time evolution from $ t_w$ to $t_w+t_*$. This is not exactly correct. That's why we say Figure \ref{earlier_than_scrambling_time_B2CORRECT} is only almost correct. To do better we'll make a connection with subregion duality.

\end{itemize}

\subsection{Relation to subregion duality}
\label{sub_region_duality}

If we want to identify the spacetime region corresponding to the uncomplexity of Bob's density matrix, it has to be inside his entanglement wedge, as it has been shown that the bulk dual of a density matrix is its entanglement wedge \cite{Dong:2016eik}.\footnote{In a talk given at the IFQ workshop on complexity at Stanford, 2017, Brian Swingle discussed various possible definitions of complexity of density matrix and identified the action of entanglement wedge as its bulk dual \cite{Swingle:2017talk}. } 

We've seen that if Bob has the subsystem B of a system AB in an entangled pure state, there are two kinds of operators he can apply: 
\begin{enumerate}
\item Relative Schmidt basis rotations between the two subsystems A and B. These can be undone from side A.
\item Unitary operations $U_B$ that cannot be undone from side A. 
\end{enumerate}

The meaning of the first kind is clear. Exactly what we mean by the second kind of operations in more general context is discussed in Appendix \ref{uncomplexity_careful_treatment}. Here, we'll restrict our discussion to the simple situation which works for a black hole, i.e. $S$ maximally entangled pairs with extra qubits added. In this context, the unitary operations $U_B$ that cannot be undone from side A are those operators affected by the extra qubits. See later discussions of Hayden-Preskill type circuit and epidemic model \cite{Hayden:2007cs}\cite{Susskind:2014jwa}\cite{Brown:2016wib}.

It has been shown in various context that the growth of the interior geometry accompanies the growth of the minimal circuit preparing a state \cite{Stanford:2014jda}\cite{Roberts:2014isa}\cite{Brown:2015lvg}\cite{Zhao:2017iul}. As discussed earlier, the first kind of operations do not belong to any single subsystem, while the second kind of operations contribute to the uncomplexity of Bob's density matrix. So it's natural to expect that the first kind of operations are responsible for the growth of the entanglement region, i.e. the spacetime region outside both entanglement wedges, and the second kind of operations are responsible for the growth of Bob's entanglement wedge.

With slight abuse of language, in what follows we'll sometimes say certain gates are stored in certain spacetime regions. What we reallly mean is the growth of certain spacetime regions accompany the growth of certain parts of the minimal circuit. 

\begin{figure}[H] 
 \begin{center}                      
      \includegraphics[width=2in]{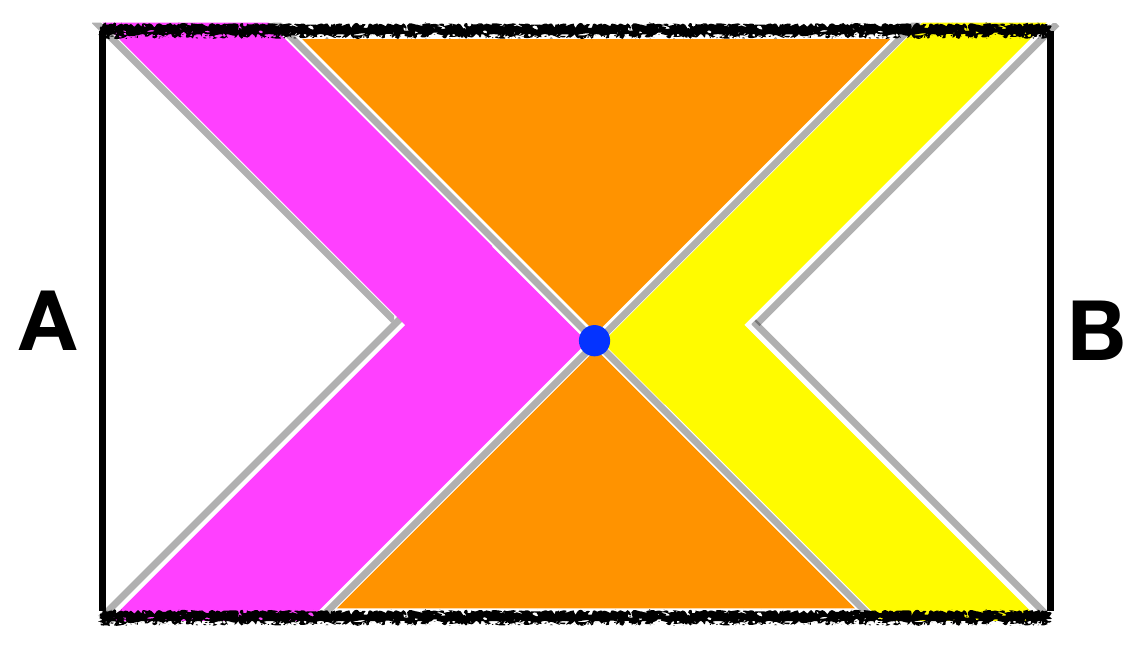}
      \caption{}
  \label{uncomplexityspacetime}
  \end{center}
\end{figure}

 In Figure \ref{uncomplexityspacetime}, the blue dot in the center represents the entangling surface, i.e. extremal surface obtained by a maximin construction \cite{Wall:2012uf}. The relative rotations of the Schmidt basis are stored in the entanglement region (orange region). This portion of spactime is accessible from both CFT's.  
 The gates of the second kind of operations are stored inside Bob's entanglement wedge (yellow region). 

To see this, we need to study the transition from two-sided to one-sided black holes in more detail. 

\subsubsection{Epidemic model}
\label{epidemic}

An epidemic model was used to study the growth of a precursor \cite{Susskind:2014jwa}\cite{Brown:2016wib}. We briefly review it here. Represent the black hole by $K$ qubits, and its dynamics by a Hayden-Preskill type circuit \cite{Hayden:2007cs}: At each time step the qubits are randomly grouped into $\frac{K}{2}$ pairs, and on each pair a randomly chosen $2$-qubit gate is applied. We can characterize the effect of some small perturbation in such a system as follows. Imagine the unpertubed system contains $K$ healthy qubits, and the perturbation is one extra qubit carrying some disease. The sick qubit enters the system at $\tau = 0$. Any qubits who interact directly or indirectly with sick qubits will get sick. We define the size of the epidemic $s(\tau)$ to be the number of sick qubits at time $\tau$. It satisfies
\begin{align*}
&\frac{ds}{d\tau} = \frac{(K+1-s)s}{K}\\
&\frac{s(\tau)}{K+1} = \frac{\frac{1}{K}e^{\frac{K+1}{K}\tau}}{1+\frac{1}{K}e^{\frac{K+1}{K}\tau}} = \frac{\frac{1}{K}e^{\tau}}{1+\frac{1}{K}e^{\tau}}
\end{align*}
We used initial condition $s(\tau = 0) = 1$. Later we'll adjust this initial condition to better match the bulk geometry.

At each time step from $\tau-\Delta\tau$ to $\tau$, $K$ gates are applied.\footnote{We count each 2-qubit gate as $2$ gates.} Among them, $s(\tau)$ gates involve the extra qubit and cannot be undone by the other side, $K-s(\tau)$ gates are healthy and can be undone by the other side. 

\subsubsection{Compare with black hole geometry}

\begin{figure}[H] 
 \begin{center}                      
      \includegraphics[width=2in]{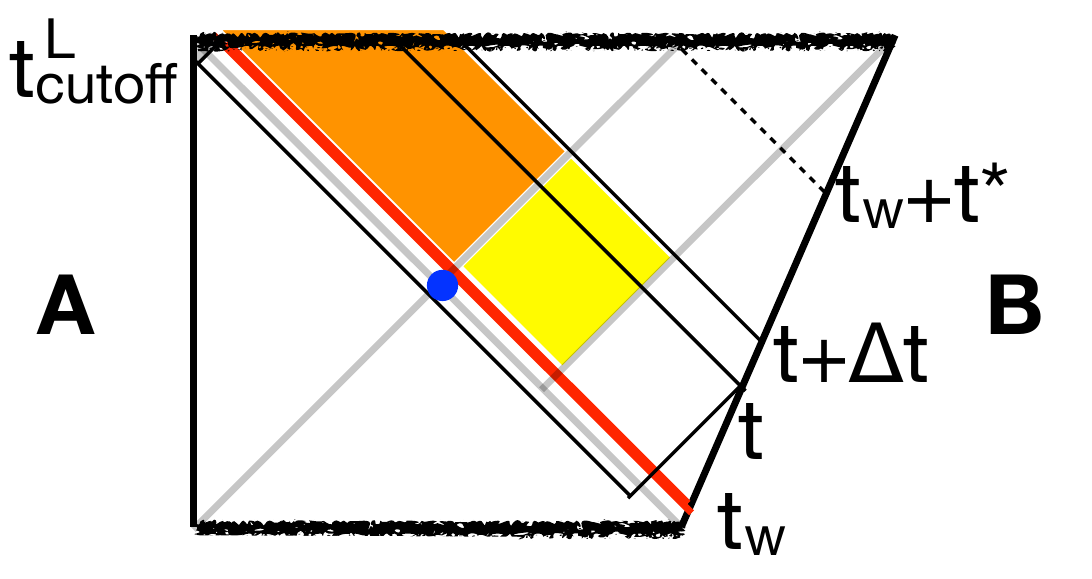}
      \caption{}
  \label{transition1}
  \end{center}
  \vspace{-0.6cm}
\end{figure}

We perturb thermofield double from Bob side at $t_w$, and focus on the regime $t_w<t<t_w+t_*$. (Figure \ref{transition1}. The blue dot is the entangling surface.)

As we increase $t$, the spacetime inside both the entanglement region (orange region in Figure \ref{transition1}) and Bob's entanglement wedge (yellow region) grows. We want to argue that the computations that can be undone by the left side are stored in the orange region (entanglement region), while the computations that cannot be undone by Alice are stored in the yellow region (Bob's entanglement wedge). We will show that the growth of these regions has close correspondence with epidemic picture.

We work in BTZ geometry, and look at the spacetime volume of these regions. We found that
\begin{align*}
&\frac{d\text{Vol}_{\text{orange}}(t)}{dt} = \pi r(t)^2 = \pi r_H^2\tanh^2\left(\frac{\pi}{\beta}(t_*-t+t_w)\right)\\
&\frac{d\text{Vol}_{\text{yellow}}(t)}{dt}   =  \frac{\pi r_H^2}{\cosh^2\left(\frac{\pi}{\beta}(t_*+t_w-t)\right)}
\end{align*}
where $t_* = \frac{\beta}{2\pi}\log S$.

To compare with the circuit picture, we first go to dimensionless time and let $\tau= \frac{\displaystyle 2\pi}{\displaystyle \beta}(t-t_w)$. Let $S = K$. Then we have
\begin{subequations}
\label{volume}
\begin{align}
&\frac{d\text{Vol}_{\text{entanglement region}}}{d\tau} = \left(\pi  r_H l^2 \right)\times
\begin{cases}
\frac{(1-\frac{1}{K}e^{\tau})^2}{(1+\frac{1}{K}e^{\tau})^2}& 0<\tau<\log K\\
0& \tau>\log K
\end{cases}\label{volume_healthy}\\
&\frac{d\text{Vol}_{\text{Bob's entanglement wedge}}}{d\tau} =\left(\pi  r_H l^2\right)\times
\begin{cases}
\frac{\frac{4}{K}e^{\tau}}{(1+\frac{1}{K}e^{\tau})^2}&0<\tau<\log K\\
1&\tau>\log K
\end{cases}\label{volume_sick}
\end{align}
\end{subequations}
where we also included the regime when $t-t_w>t_*$. 

To do the match, we shift the circuit time by $\log 4$. The circuit answer becomes
\begin{subequations}
\label{circuit}
\begin{align}
&\frac{dN_{\text{can be undone}}(\tau)}{d\tau} =  K-s(\tau) = K \frac{1}{1+\frac{4}{K}e^{\tau}}\label{qubit_healthy}\\
&\frac{dN_{\text{cannot be undone}}(\tau)}{d\tau} = s(\tau) = K \frac{\frac{4}{K}e^{\tau}}{1+\frac{4}{K}e^{\tau}}\label{qubit_sick}
\end{align}
\end{subequations}
To compare \eqref{volume} and \eqref{circuit}, we only need to compare \eqref{volume_sick} and \eqref{qubit_sick}. We identify spacetime volume $\pi  r_H l^2$ in \eqref{volume} with entropy $K$.\footnote{  $\frac{\displaystyle\pi r_H l^2}{\displaystyle G_N l^2} = 2K$}
The following plots are made for two scrambling times after the perturbation. 

We first take the value of $K$ to be $10^5$, with which the scrambling time $\tau_* = \log K$ is roughly 10.

 \begin{figure}[H]
 \begin{center}
  \begin{subfigure}[b]{0.4\textwidth}
  \begin{center}
    \includegraphics[scale=0.5]{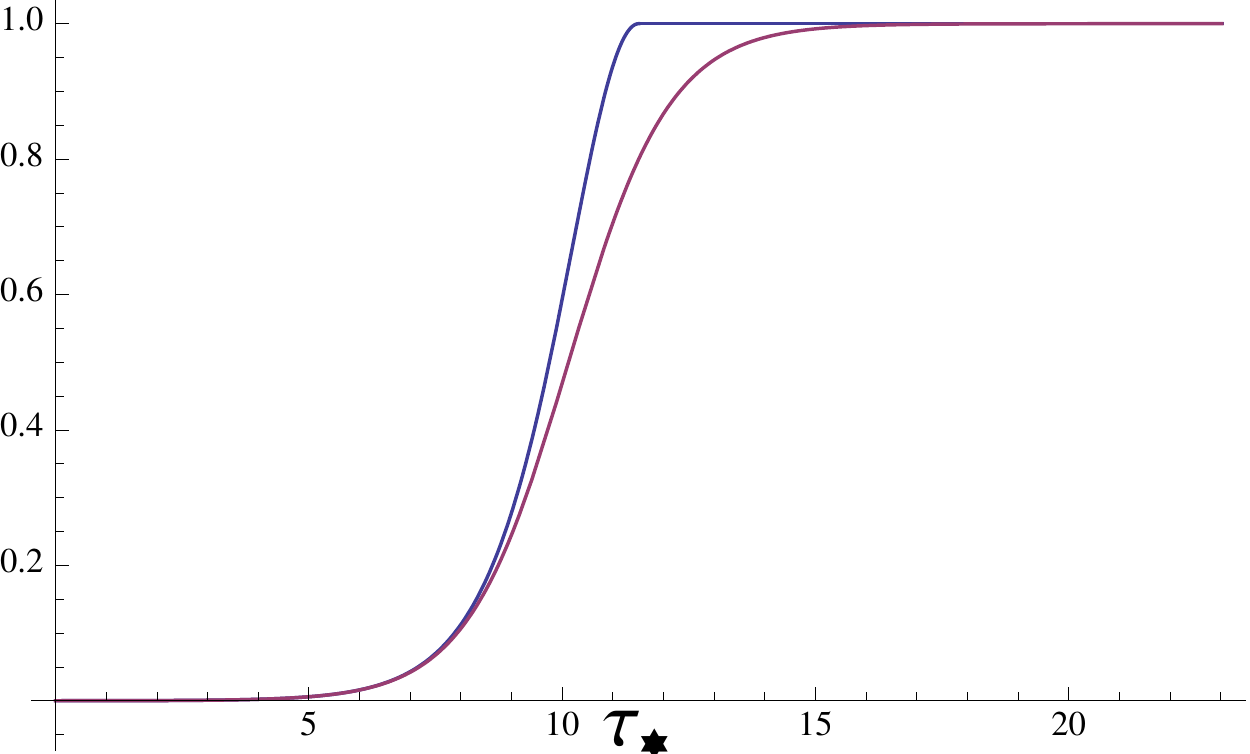}
    \caption{The growth of Bob's entanglement wedge and the growth of Bob's gates that cannot be undone by Alice.}
    \label{sick_compare2}
    \end{center}
  \end{subfigure}
  \hspace{2em}
  \begin{subfigure}[b]{0.4\textwidth}
  \begin{center}
    \includegraphics[scale=0.5]{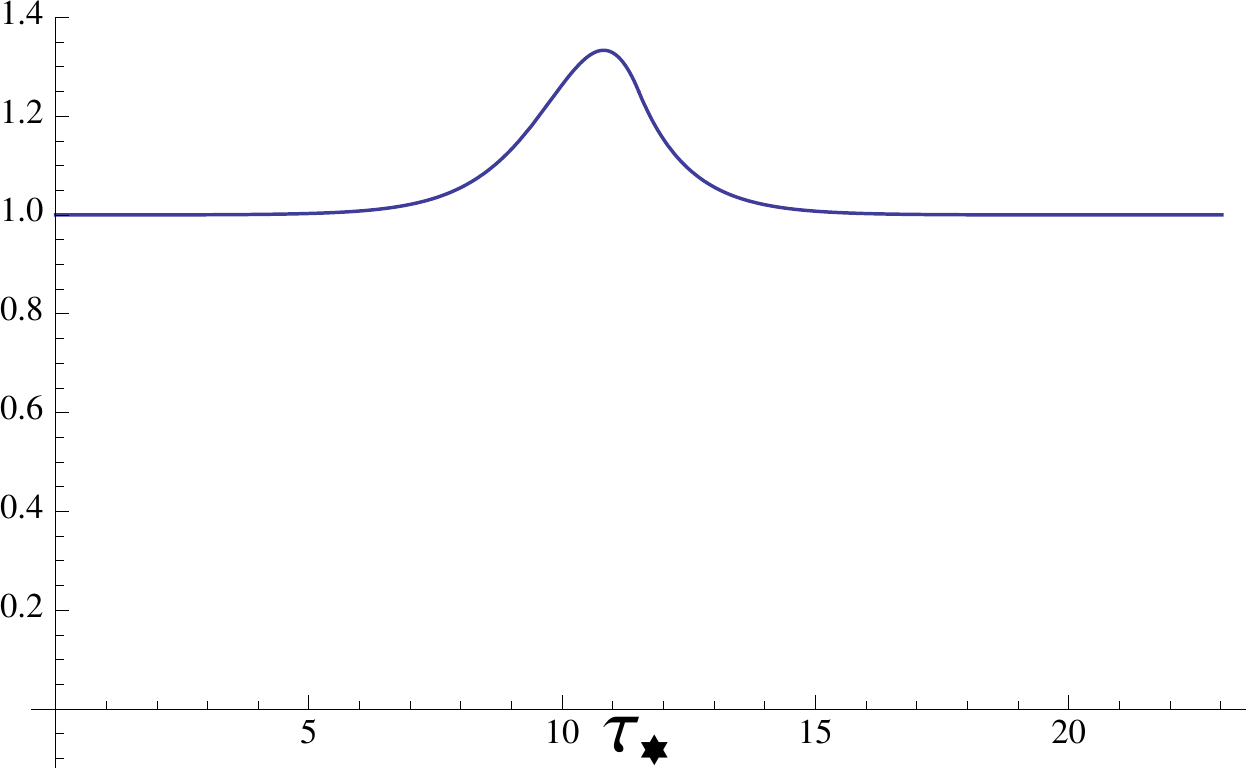}
    \caption{The ratio of the two results in (a).}
    \label{sick_ratio2}
    \end{center}
  \end{subfigure}
  \caption{$\log K = 11$}
  \vspace{-.5cm}
  \label{compare2}
  \end{center}
\end{figure}

In Figure \ref{sick_compare2}, the blue line is the increase of spacetime volume in Bob's entanglement wedge every thermal time (Equation \eqref{volume_sick}). The pink line is the increase of the number of gates that cannot be undone from Alice side (Equation \eqref{qubit_sick}). Figure \ref{sick_ratio2} is the ratio of these two. We see that the results from black hole spacetime volume calculation deviate from the results from circuit picture analysis for roughly $4$ thermal times around the scambling time. However, as we increase the number of qubits (increase scrambling time), a couple of thermal times become less and less significant, and the match becomes better and better. In the following plots we take $K$ to be the entropy of a solar mass black hole, $10^{76}$, with which $\log K = 175$.

\begin{figure}[H]
 \begin{center}
  \begin{subfigure}[b]{0.4\textwidth}
  \begin{center}
    \includegraphics[scale=0.5]{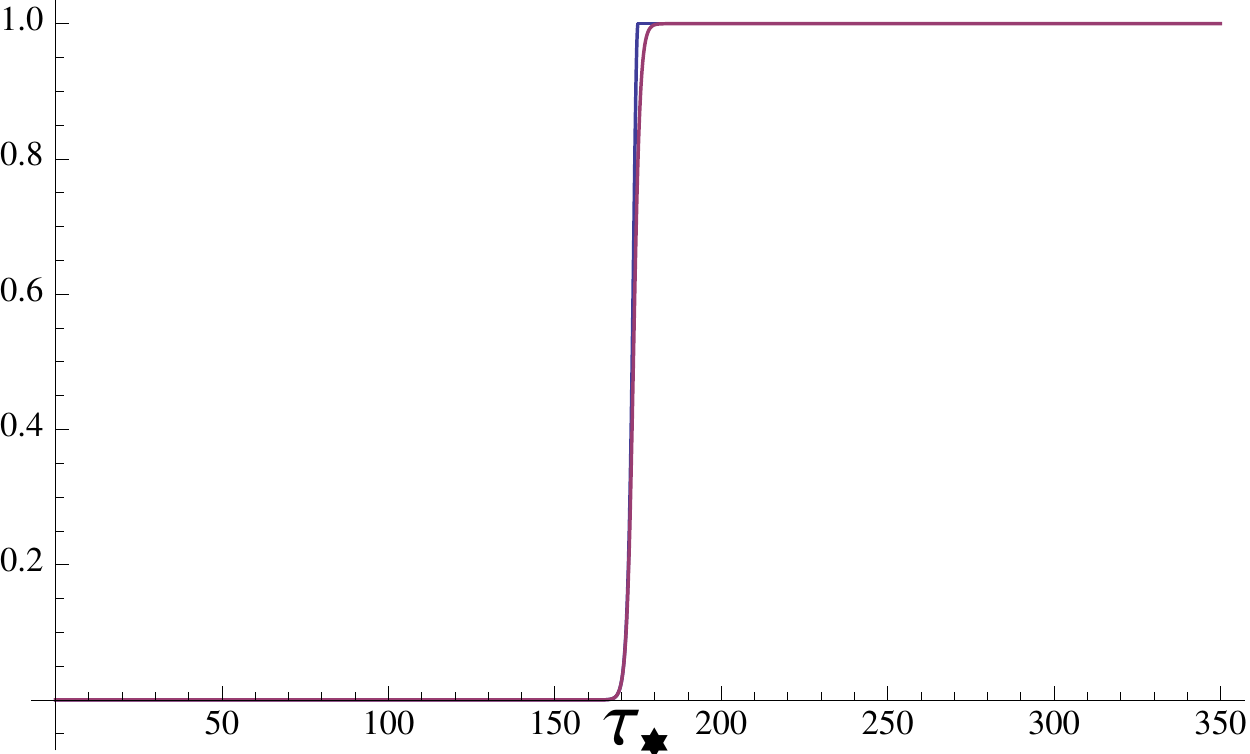}
    \caption{The growth of Bob's entanglement wedge and the growth of Bob's gates that cannot be undone by Alice.}
    \label{sick_compare2}
    \end{center}
  \end{subfigure}
  \hspace{2em}
  \begin{subfigure}[b]{0.4\textwidth}
  \begin{center}
    \includegraphics[scale=0.5]{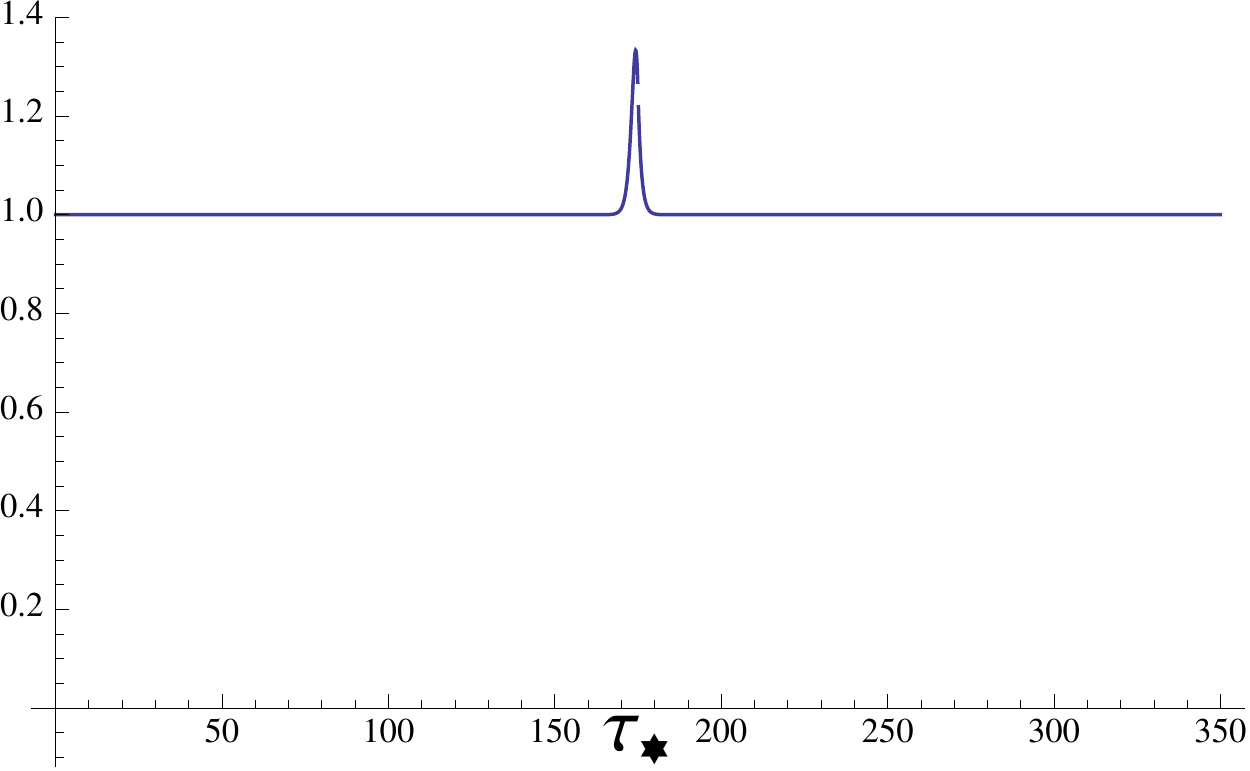}
    \caption{The ratio of the two results in (a).}
    \label{sick_ratio2}
    \end{center}
  \end{subfigure}
  \caption{$\log K = 175$}
  \vspace{-.5cm}
  \label{compare2}
  \end{center}
\end{figure}

We see that aside from a few thermal times around the scrambling time, at which the bulk geometry involved is close to the singularity, the match is almost perfect. Note that the functional forms of \eqref{volume} and \eqref{circuit} are not the same. We also don't expect them to be the same. After all, the epidemic model is a toy model, and we don't expect the spacetime volume to exactly reproduce the gates counting. What this match tells us is that the spacetime growth (both inside and outside Bob's entanglement wedge) and the qubit model share the same key features like exponential growth, saturation after scrambling time, and so on. So it is reasonable to identify different parts of the circuits as being stored in different spacetime regions. \\

Coming back to uncomplexity, the computational power of Bob's density matrix comes from Bob's operations that cannot be undone from Alice side. We see that in a black hole geometry those gates are stored in Bob's entanglement wedge. So Bob's uncomplexity in general should not correspond to the entire interior region accessible to him. Only the portion which is also in his entanglement wedge contributes to his uncomplexity. This solves the puzzle about the uncomplexity of thermofield double we encountered in section \ref{simple_examples}.
 
Next we'll look at this this from our definition \eqref{uncomplexitydef}.

\subsubsection{Uncomplexity during transition from two-sided to one-sided black holes, revisited}

In section \ref{transition} we made some approximations in studying Bob's uncomplexity during less than scrambling time after the perturbation. Now with a better understanding of different gates being stored in different parts of spacetime, we can do better. 

\begin{figure}[H] 
 \begin{center}                      
      \includegraphics[width=5in]{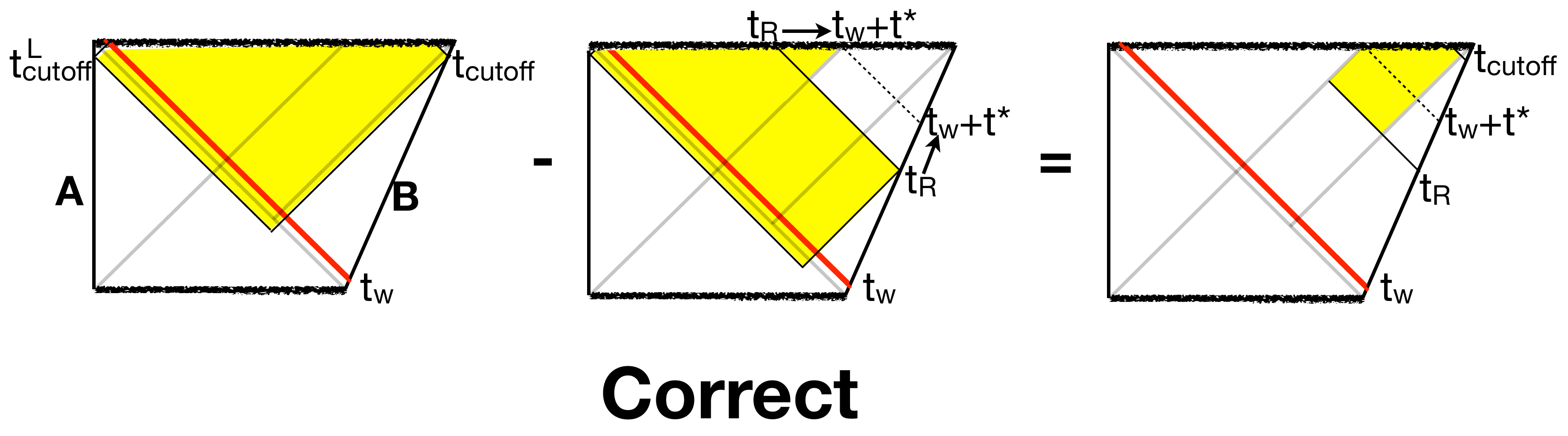}
      \caption{}
  \label{earlier_than_scrambling_time_B2CORRECT2}
  \end{center}
\end{figure}

Consider the case when we fix the left cutoff time at some large positive value, $t_w<t_R<t_w+t_*$. As argued before, in maximizing the second term in \eqref{uncomplexitydef} we could do some right time evolution without changing Bob's density matrix. But only those relative Schmidt basis rotations will leave Bob's density matrix truly untouched. So as we evolve up we can only apply those gates stored in the entanglement region (middle Penrose diagram in Figure \ref{earlier_than_scrambling_time_B2CORRECT2}). At the end, what corresponds to Bob's uncomplexity is exactly the interior spacetime accessible to him which is also inside his entanglement wedge (right Penrose digram in Figure \ref{earlier_than_scrambling_time_B2CORRECT2}).

Similar argument can be made if we fix the left cutoff time at some large negative value and we will reach the same conclusion. The details are in Appendix \ref{othercutoff}.

\subsection{Apparent horizons}

\begin{figure}[H] 
 \begin{center}                      
      \includegraphics[width=2.6in]{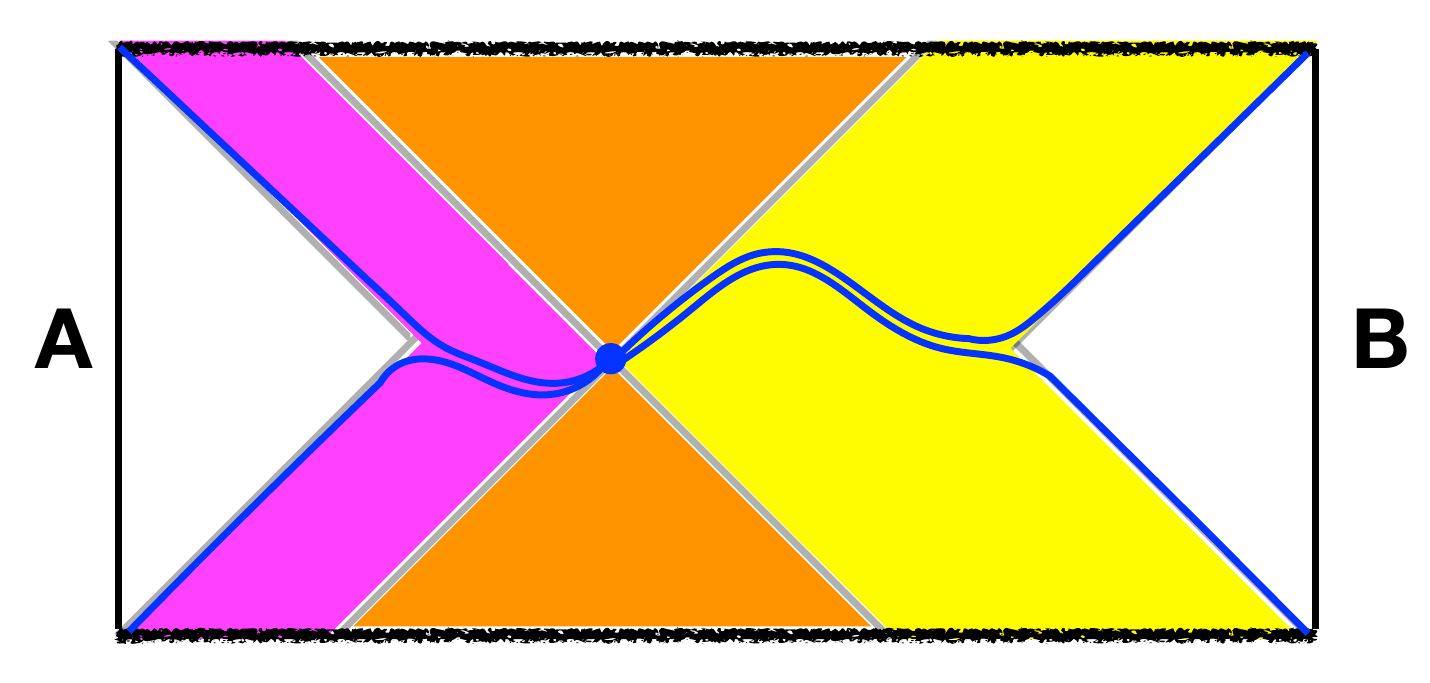}
      \caption{}
  \label{general_Penrose}
  \end{center}
  \vspace{-0.6cm}
\end{figure}

Here is the general picture we have so far. Consider a wide Penrose diagram for which the entangling surface lies behind the event horizon. (Figure \ref{general_Penrose}. The Blue dot in the center is the entangling surface.) We again focus on side B. Notice that the area of the event horizon is bigger than the area of the entangling surface, which means the coarse-grained entropy $S_{\text{coarse-grained}}$ is bigger than the entanglement entropy $S_{\text{ent}}$. 

\begin{figure}[H] 
 \begin{center}                      
      \includegraphics[width=3.4in]{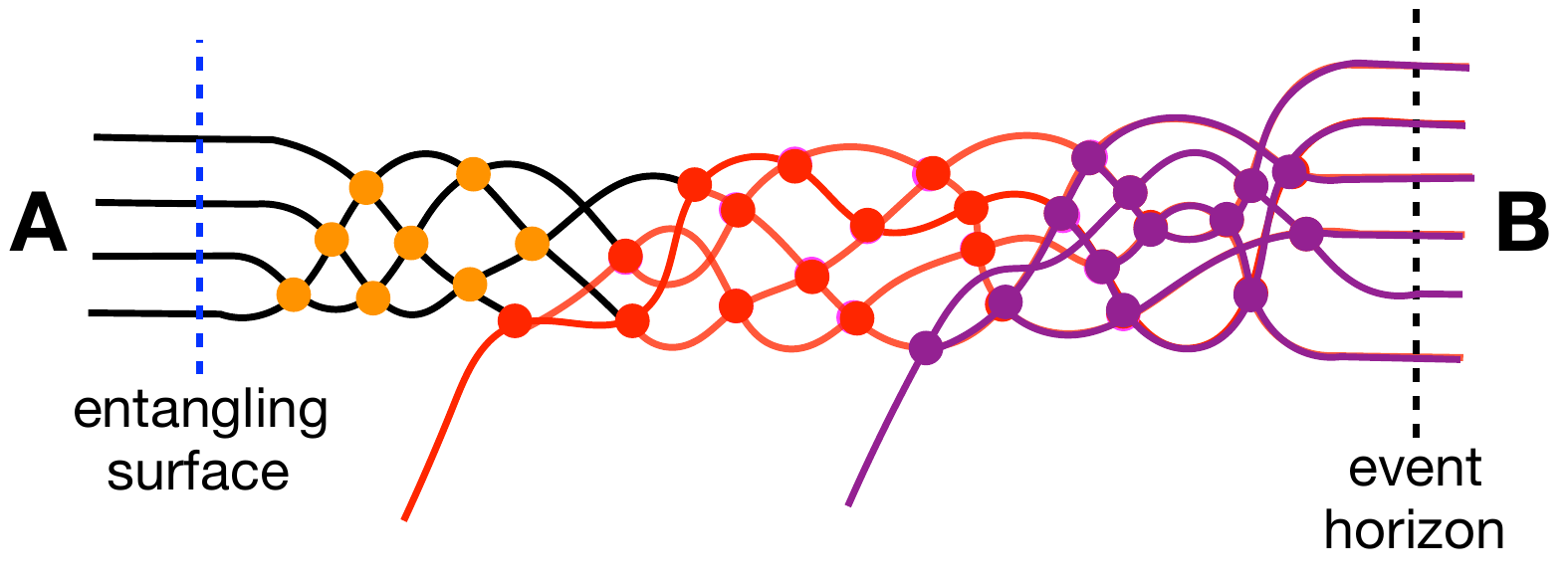}
      \caption{}
  \label{general_circuit}
  \end{center}
  \vspace{-0.6cm}
\end{figure}

We start from the entangling surface whose area gives the entanglement entropy, and move towards the event horizon along a spacelike direction. Here is a circuit description. At the entangling surface there are $S_{\text{ent}}$ qubits actively computing. As we move towards the event horizon there are more and more qubits coming and actively participating in the computation. (See Figure \ref{general_circuit}.) This is reflected in the fact that the coarse-grained entropy increases.

\begin{figure}[H] 
 \begin{center}                      
      \includegraphics[width=3.4in]{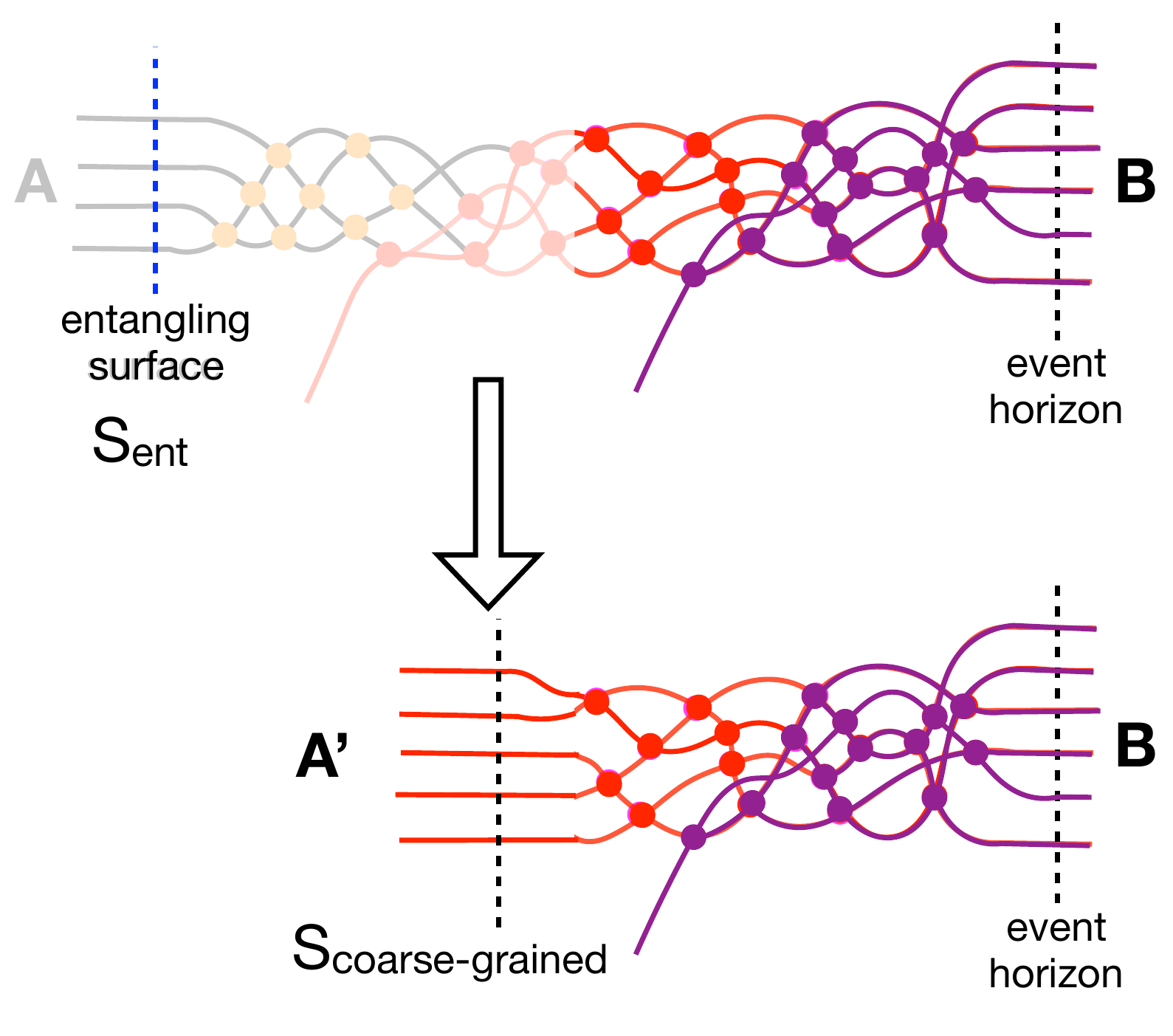}
      \caption{}
  \label{general_circuit_coarse_graining}
  \end{center}
  \vspace{-0.6cm}
\end{figure}

Recently it was shown that the area of the apparent horizon is associated with the coarsed-grained entropy subject to knowing the exterior geometry \cite{Engelhardt:2017aux}. It is natural to expect the coarsed-grained entropy given by the area of the apparent horizon corresponds to the number of qubits actively computing whose gates are stored in the nearby spacetime region. In fact, from a circuit point of view, the procedure of coarse graining over the information in the interior exactly corresponds to the procedure of erasing some left part of the circuit in Figure \ref{general_circuit} and replace it with $S_{\text{coarse-grained}}$ bell pairs with another system $A'$ (Figure \ref{general_circuit_coarse_graining}).\footnote{I thank Aron Wall for explaining this to me.}

Notice that the apparent horizon passes the entangling surface (blue lines in Figure \ref{general_Penrose}). Now we can describe our general picture as follows. For a wide Penrose diagram, at the entangling surface there are $S_{\text{ent}}$ Bell pairs whose relative rotations are stored inside the entanglement region (orange region in Figure \ref{general_Penrose}). As we move away from the entangling surface along the apparent horizon, the coarse-grained entropy is increasing. Correspondingly, more and more quibits participate in the computation. The number of qubits actively computing is given by the apparent horizon area $S_{\text{coarse-grained}}$. The gates that cannot be undone from Alice side are stored in Bob's entanglement wedge. 

\section*{Acknowledgements}
I thank Adam Brown,  Aron Wall for discussions. I am especially grateful to Leonard Susskind for discussions, for pushing me to sharpen the statement, and for a lot of encouragement. 

\appendix

\section{A more careful treatment of the definition of uncomplexity}
\label{uncomplexity_careful_treatment}

\begin{figure}[H] 
 \begin{center}                      
      \includegraphics[width=2in]{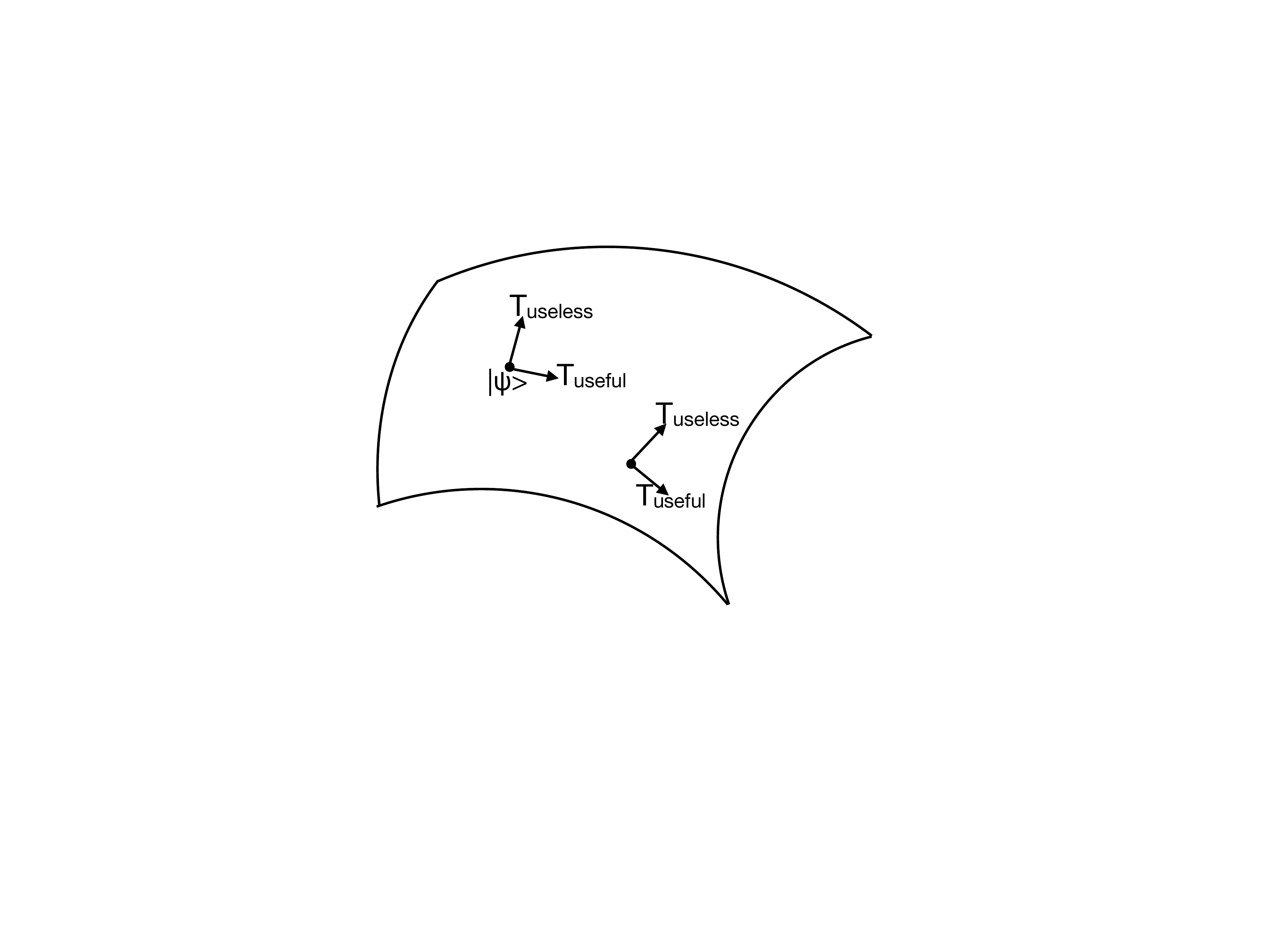}
      \caption{The Hilbert space $\mathcal{H}_A\otimes\mathcal{H}_B$ equipped with complexity metric.}
  \label{complexity_metric_tangent_space}
  \end{center}
  \vspace{-0.6cm}
\end{figure}

The key to give a proper definition of mixed state uncomplexity is to get rid of those relative Schmidt basis rotations, as Bob cannot do computations with them. Bob's Schmidt basis rotations depend on the state. Consider the space of all states in the Hilbert space $\mathcal{H}_A\otimes\mathcal{H}_B$ equipped with complexity metric \cite{2007quant.ph..1004D}\cite{Brown:2016wib}\cite{Brown:2017jil}. At each point $\ket{\psi}$ on this space, Bob can apply unitary operators $U_B$ to move the state $\ket{\psi}$ in different directions. The relative Schmidt basis rotations form a subspace of the tangent space at each point, which we call $\mathcal{T}_{\text{useless}}$. The tangent vectors orthogonal to the relative Schmidt basis rotations form another subspace which we call $\mathcal{T}_{\text{useful}}$.\footnote{The words useless and useful are used because Bob can not do computations with the former, while he can do computations with the latter operations.} (Figure \ref{complexity_metric_tangent_space}) 

When we ask about Bob's uncomplexity, we want to ask, starting from $\ket{\psi}$, how far can Bob go if we only allow his unitaries to move along those useful directions? In definition \eqref{uncomplexitydef}, what we do is to first let Bob move along the useless directions for as far as possible. Say, Bob stops at $\ket{\psi'}$. Now set $\ket{\psi'}$ as his new starting point, and ask how far he can go. Assuming how far he can go along useful directions does not depend on the purifications he starts from (Uncomplexity of a density matrix does not depend on its purifications), \eqref{uncomplexitydef} gives Bob's computational power.

\begin{figure}[H] 
 \begin{center}                      
      \includegraphics[width=2.6in]{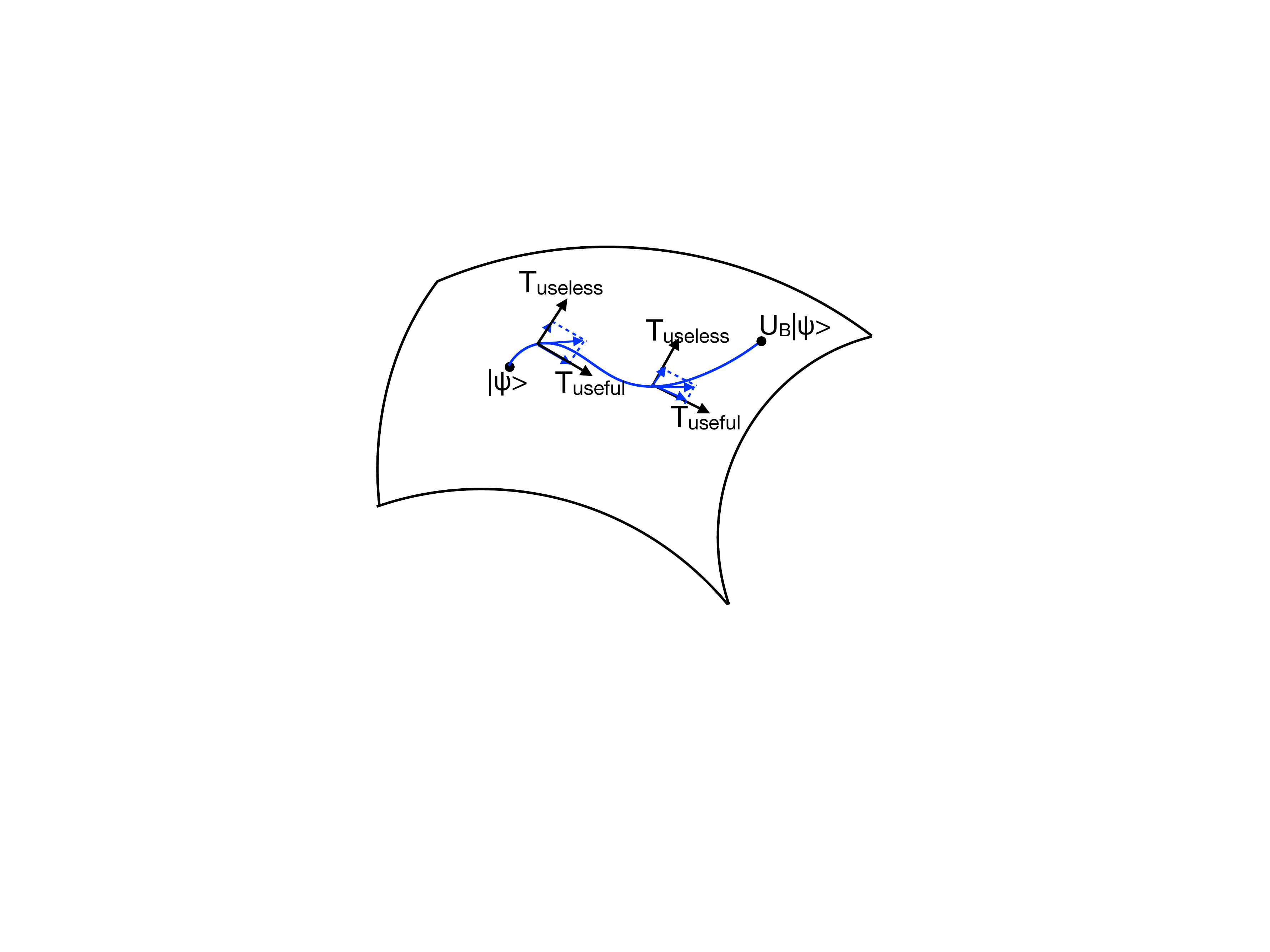}
      \caption{A minimal circuit connecting $\ket{\psi}$ and $U_B\ket{\psi}$.}
  \label{complexity_metric_tangent_space_path}
  \end{center}
  \vspace{-0.6cm}
\end{figure}

We classify Bob's unitary operations into two kinds. The first kind are the $U_B$'s that can be undone by Alice. The second kind are the $U_B$'s that cannot be undone by Alice. In section \ref{sub_region_duality} we argue that the first kind of gates are stored in the entanglement region, while the second of gates are stored in Bob's entanglement wedge. Here is what we mean. Consider the minimal circuit from $\ket{\psi}$ to $U_B\ket{\psi}$ (blue line in Figure \ref{complexity_metric_tangent_space_path}). We decompose each small step of the circuit into $\mathcal{T}_{\text{useless}}$ and $\mathcal{T}_{\text{useful}}$ directions. The gates in $\mathcal{T}_{\text{useless}}$ directions are relative rotations of the Schmidt basis and are stored in the entanglement region. The gates along $\mathcal{T}_{\text{useful}}$ directions cannot be undone by Alice and are stored in Bob's entanglement wedge.

\section{Calculations with large negative left cutoff time}
\label{othercutoff}

Consider thermofield double perturbed by Bob throwing in a thermal photon at time $t_w$. We will calculate Bob's uncomplexity with the left cutoff time fixed at some large negative value. 

\begin{figure}[H] 
 \begin{center}                      
      \includegraphics[width=5in]{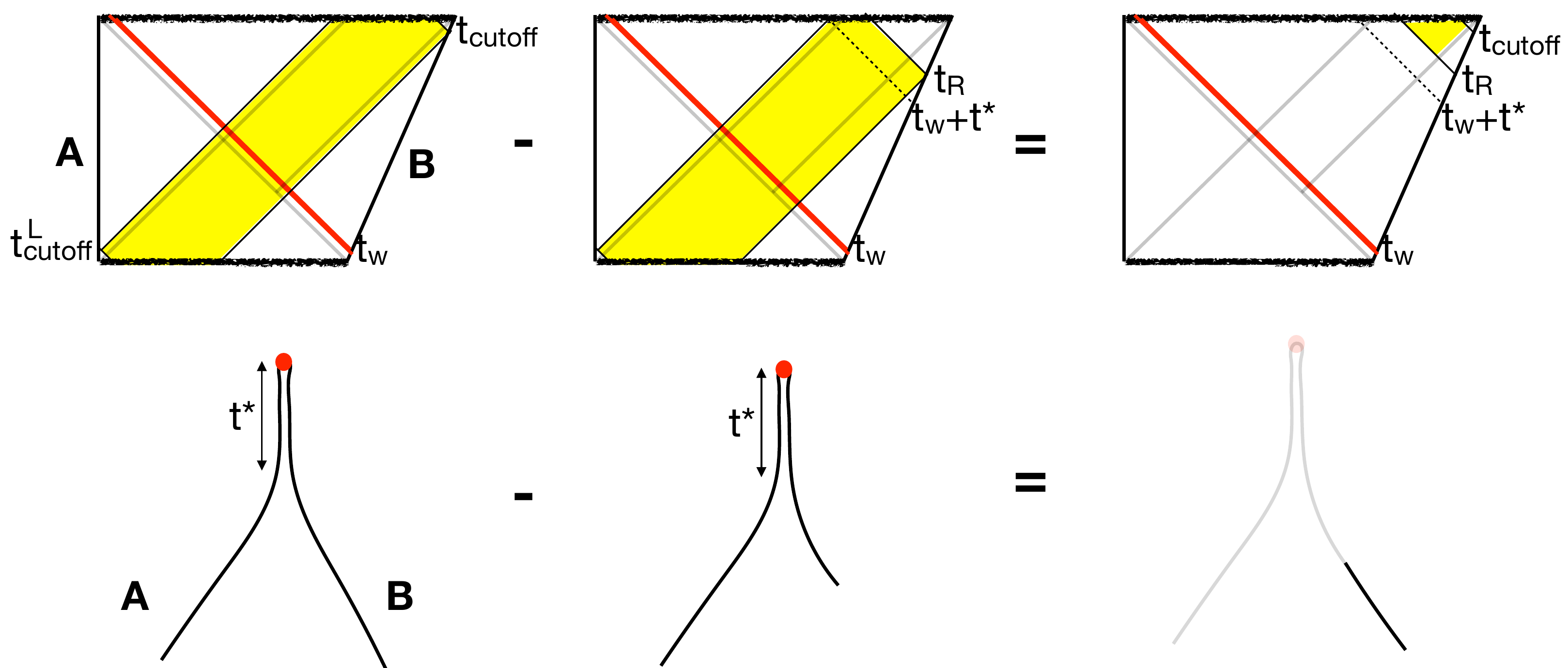}
      \caption{}
  \label{later_than_scrambling_time_B}
  \end{center}
\end{figure}

For $t_R>t_w+t_*$, everything stays the same as the case of large positive left cutoff time (Figure \ref{later_than_scrambling_time_B}). This is expected, since in this regime Bob has a one-sided black hole whose expansion is fueled by his own uncomplexity. Nothing about his black hole should be affected by the value of the left time. 

\begin{figure}[H] 
 \begin{center}                      
      \includegraphics[width=5in]{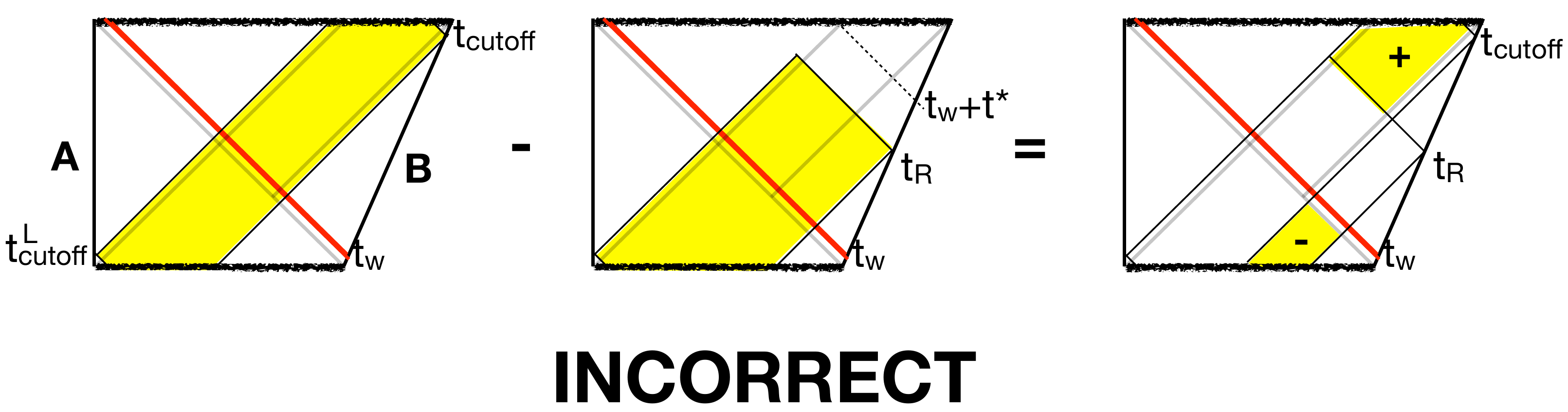}
      \caption{}
  \label{earlier_than_scrambling_time_BINCORRECT}
  \end{center}
\end{figure}

The Penrose diagrams look different when $t_w<t_R<t_w+t_*$. Naively Bob's uncomplexity starts to decrease as we further decrease the right time beyond $t_w+t_*$ (Figure \ref{earlier_than_scrambling_time_BINCORRECT}. Notice the two regions in the right Penrose diagram contribute with different signs.). But this is again incorrect. 

\begin{figure}[H] 
 \begin{center}                      
      \includegraphics[width=5in]{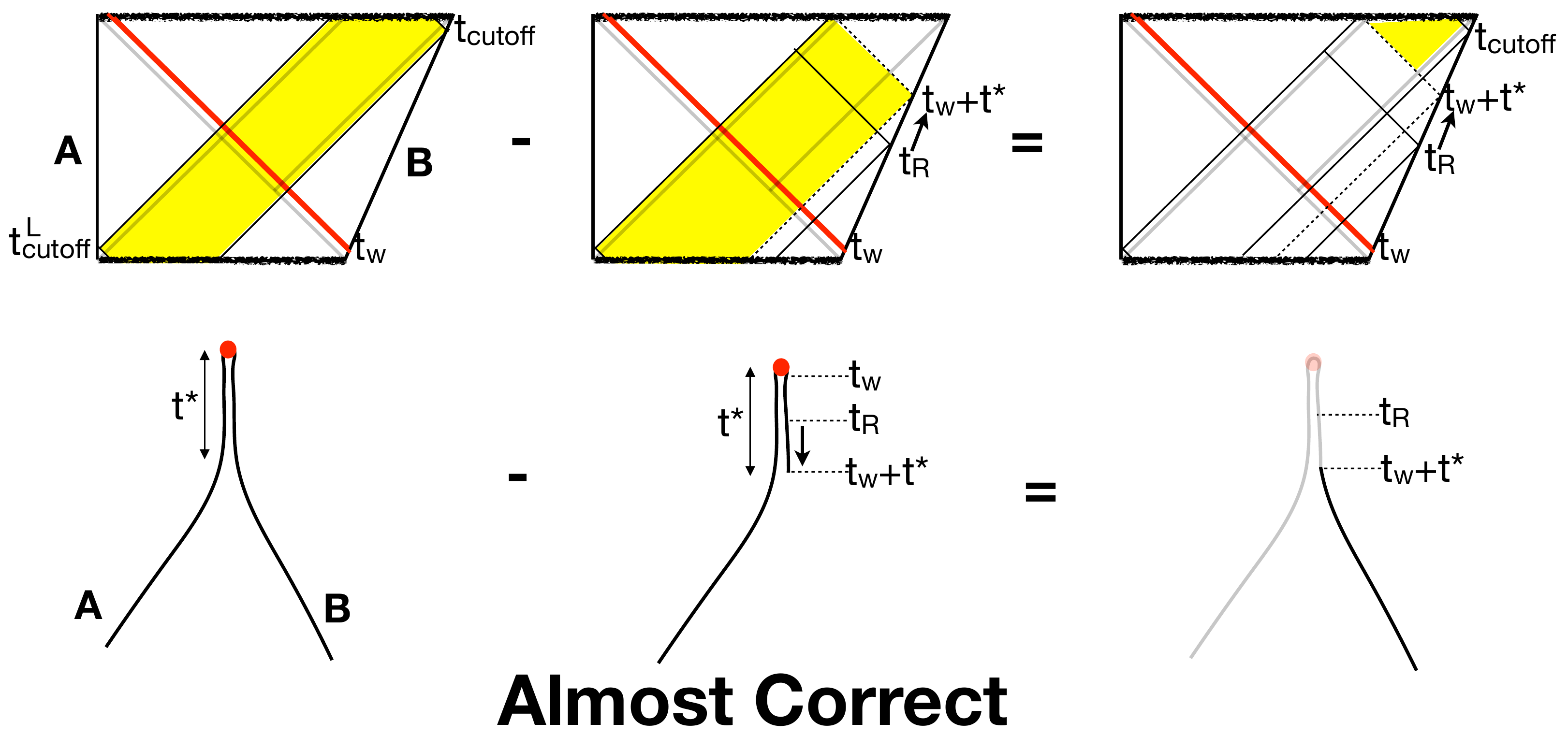}
      \caption{}
  \label{earlier_than_scrambling_time_BCORRECT}
  \end{center}
\end{figure}

As before, within less than scrambling time after the perturbation, Bob is in the two-sided black hole regime. With fixed left time at some large negative cutoff, if we move the right time (Bob's time) below $t_w+t_*$, the total wormhole length will exceed the cutoff length. We need to either move the left cutoff time, or equivalently, move the right time back to $t_w+t_*$ (Recall that this will not change Bob's density matrix.) Here is an almost correct picture (Figure \ref{earlier_than_scrambling_time_BCORRECT}), and we conclude that Bob's uncomplexity stays constant when $t_w<t_R<t_w+t_*$. 

This is not exactly correct. In section \ref{sub_region_duality} we see that the relative Schmidt basis rotations are stored in the entanglement region, while Bob's gates that cannot be undone by Alice are stored in Bob's entanglement wedge. To get Bob's uncomplexity completely correct we need to use this conclusion. 

\begin{figure}[H] 
 \begin{center}                      
      \includegraphics[width=5in]{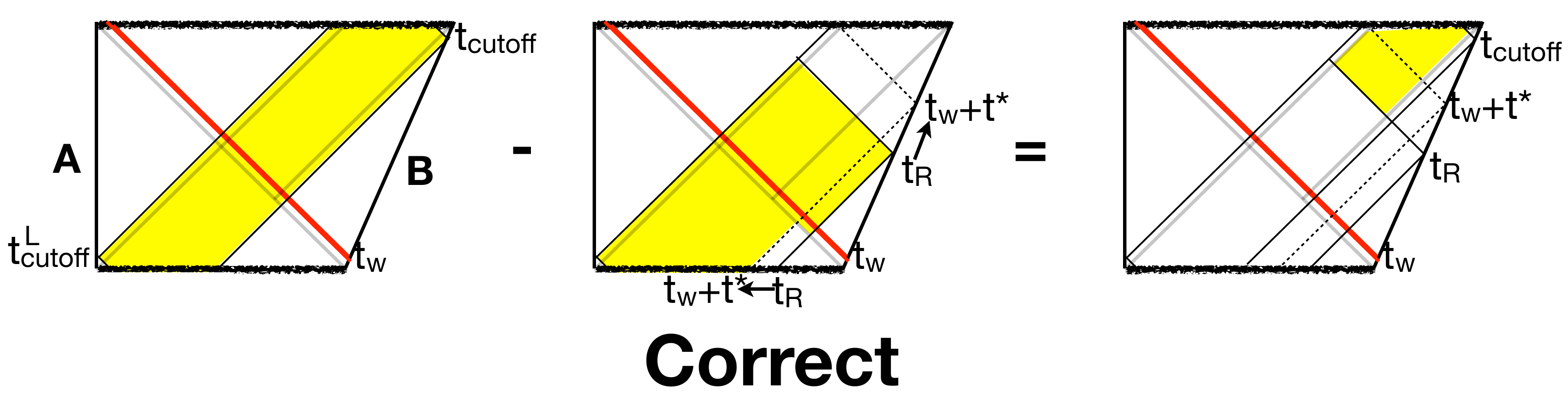}
      \caption{}
  \label{earlier_than_scrambling_time_BCORRECT2}
  \end{center}
\end{figure}

We look at Bob's uncomplexity at time $t_R<t_w+t_*$. To avoid exceeding the maximal cutoff length in the second term of \eqref{uncomplexitydef}, we need to do right time evolution until $t_w+t_*$. But to make sure Bob's density matrix does not change we can only apply those gates stored inside the entanglement region (behind the past horizon, see the middle Penrose diagram in Figure \ref{earlier_than_scrambling_time_BCORRECT2}). At the end we again see that Bob's uncomplexity corresponds to his accessible interior which is also inside his entanglement wedge (right Penrose diagram in Figure \ref{earlier_than_scrambling_time_BCORRECT2}). 

\bibliographystyle{unsrt}

\bibliography{reference}

\end{document}